\documentclass[a4paper,11pt]{article}
\pdfoutput=1
cm \voffset = -2truecm

\usepackage{graphicx}
\usepackage{amsmath}
\usepackage{amssymb}
\usepackage{latexsym}
\usepackage[colorlinks]{hyperref}
\usepackage{color}
\usepackage{cite}
\usepackage{makeidx}
\usepackage{float}

\newcommand{\bra}{\begin{array}}
\newcommand{\era}{\end{array}}
\newcommand{\beq}{\begin{equation}}
\newcommand{\eeq}{\end{equation}}
\newcommand{\bqr}{\begin{eqnarray}}
\newcommand{\eqr}{\end{eqnarray}}

\def\BC{\bb C}
\def\_\BC{\bbi C}

\def\no2 {{\textstyle{n\over 2}}}

\newcommand{\lb}{\label}

\begin{document}
\begin{titlepage}
\setcounter{page}{1}
\renewcommand{\thefootnote}{\fnsymbol{footnote}}

\begin{flushright}
\end{flushright}

\vspace{5mm}
\begin{center}

{\Large \bf {
Goos-H\"{a}nchen Shifts in
AA-Stacked Bilayer Graphene Superlattices}}

\vspace{5mm}

{\bf Youness Zahidi}$^{a}$, {\bf Ilham Redouani}$^{a}$ and {\bf
Ahmed Jellal\footnote{\sf ajellal@ictp.it --
a.jellal@ucd.ac.ma}}$^{a,b}$

\vspace{5mm}

{$^{a}$\em Theoretical Physics Group,  
Faculty of Sciences, Choua\"ib Doukkali University},\\
{\em PO Box 20, 24000 El Jadida, Morocco}

{$^b$\em Saudi Center for Theoretical Physics, Dhahran, Saudi
Arabia}






\vspace{3cm}

\begin{abstract}

The quantum Goos-H\"{a}nchen shifts of the transmitted electron
beam through an AA-stacked bilayer graphene superlattices is
investigated.
We found that the band structures of graphene
superlattices can have more than one Dirac point, their locations
do not depend on the number of barriers.
It was revealed that any $n$-barrier structure is perfectly
transparent at normal incidence around the Dirac points created in
the superlattices. We showed
that the Goos-H\"{a}nchen shifts display sharp peaks inside the
transmission gap around  two Dirac points
($E= V_B + \tau$, $E= V_W + \tau$), which are equal to those of
transmission resonances. 
The obtained Goos-H\"{a}nchen shifts are exhibiting negative as well as positive behaviors
and strongly depending on the location of Dirac points.
It is observed that the maximum
absolute values of the shifts increase as long as
the number of barriers is increased.
Our analysis is done by considering four cases: single, double barriers, superlattices without and with
defect.






\end{abstract}
\end{center}

\vspace{3cm}

\noindent PACS numbers: 73.22.Pr, 72.80.Vp, 71.10.Pm, 03.65.Pm\\
 \noindent Keywords:
AA-stacked bilayer graphene, barriers, superlattices, transmission,
Goos-H\"{a}nchen shifts.

\end{titlepage}


\section{Introduction}

\begin{figure}[h!]
\centering
\includegraphics[width=6cm, height=5cm]{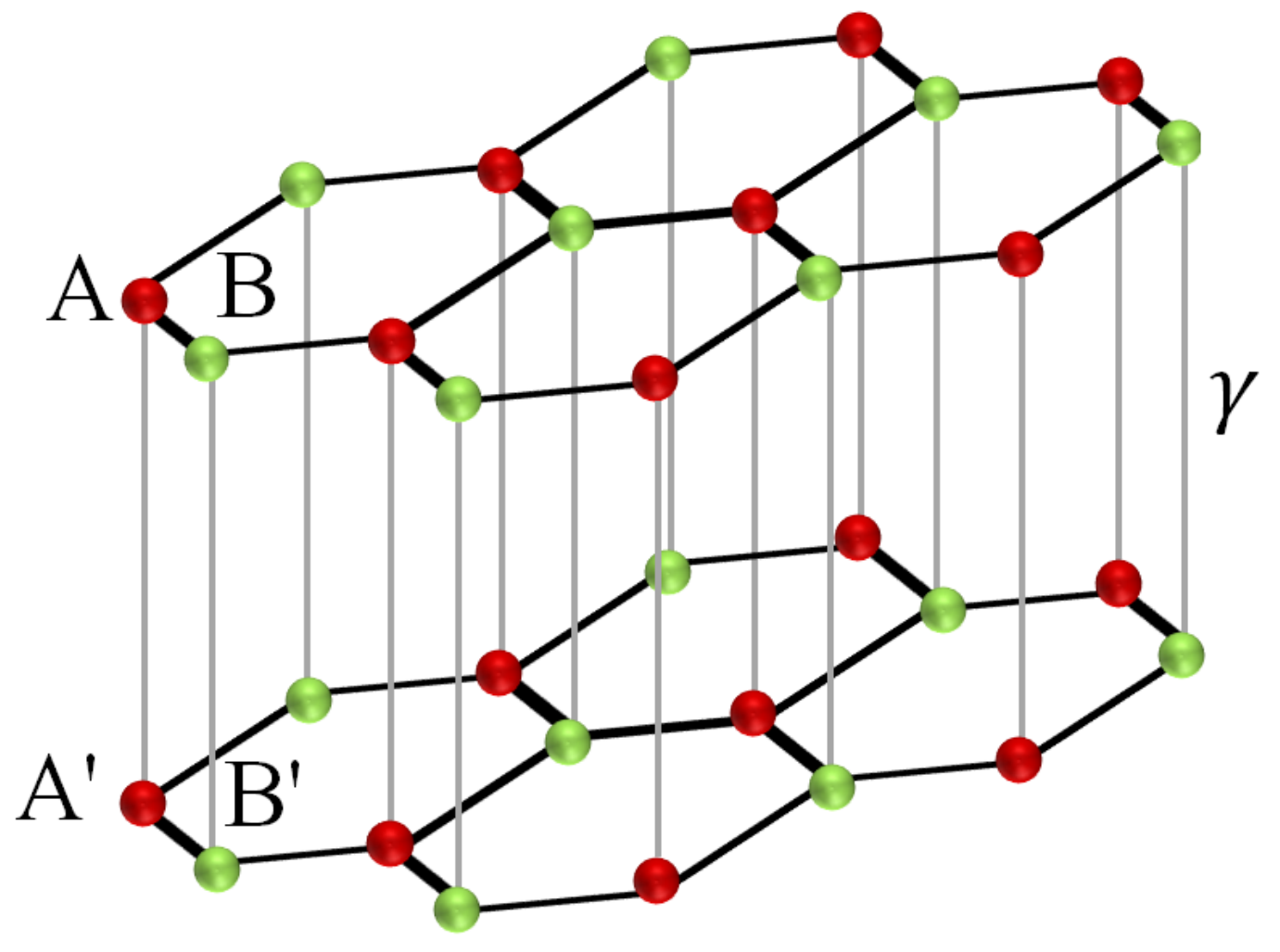}
\caption{ {\sf {Schematic illustration of lattice structure of
AA-stacked bilayer graphene, which consists of two graphene
layers. Each carbon atom of the upper layer is located above the
corresponding atom of the lower layer and they are separated by an
interlayer coupling energy $\gamma$. The unit cell of the
AA-stacked bilayer graphene consists of four atoms A, B, A' and
B'.}}}\label{struc}
\end{figure}

Since the first experimental fabrication of monolayer
graphene~\cite{NGMJ.04}, a single sheet of carbon honeycomb, it
inspired researchers due to its unique electronic properties. This
new material has a number of interesting properties, which makes
it one of the most promising materials for future
nanoelectronics~\cite{GN07}. What makes graphene so attractive
is its band structure, which is gapless and exhibits a linear
dispersion relation at two inequivalent points ($K$, $K'$) in the
vicinity of the Fermi energy. Moreover, its low energy of
electrons is governed by a (2+1) dimensional Dirac equation, witch
leads to many fascinating physical properties,
such as
Klein tunneling~\cite{KNG06}. Graphene can not only exist in the
free state, but two or more layers can stack above each other to
form what is called few layer graphene,
as the case for bilayer graphene (BLG), two stacked sheets. There
are two dominant ways in which the two layers can be stacked to
form AB or AA, with A and B are two sublattices of each layer, see
Figure~\ref{struc} for AA.

Owing to the interlayer interactions between the two layers and
stacking sequence, the energy bands of AA-stacked BLG differ from
those of the monolayer graphene and AB-stacked BLG. There are two
pairs of linear bands intersecting at the Fermi level, that are a
double copies of single layer graphene bands shifted up and down
by the interlayer coupling $\gamma\approx0.2\ eV$~\cite{Tabert12}.
Due to this special band structure, the AA-stacked BLG shows many
interesting properties~\cite{Tabert12,T11,HG11,PSBF11,RRSN12,BF13}
that are different from those of others.
The research was less focused on AA-stacked BLG then AB-stacked
BLG because of its instability, but actually recent experiment
proved that one can produce a stable AA system
~\cite{LSHL09,BSP11,ARV08}.

On the other hand, the Goos-H\"{a}nchen (GH) shift~\cite{GH47} is
a phenomenon that originated in classical optics in which a light
beam reflecting off a surface is spatially shifted as if it had
briefly penetrated the surface before bouncing back. The GH shift
was discovered by Hermann Fritz Gustav Goos and Hilda
H\"{a}nchen~\cite{GH47,GH49} and theoretically explained by
Artman~\cite{A49} in the late of 1940s. Usually, the absorption
and transmission of the two optical materials must be weak enough
to allow a reflected beam to be formed. Since its discovery, the
beam spatial shift at total reflection suspected by Newton's
corpuscular theory has been extended to other fields of physics,
such as quantum mechanics, plasma physics,
 acoustics~\cite{BLS00}, metamaterial~\cite{WZ10}, neutron
physics~\cite{HPR.10} and graphene~\cite{BSAT09,JRZH14,JWZM15}.
The Quantum version of the GH shifts is an analogue to the optical GH
one, which is referred to a lateral shift between the reflected
and incident beams occurring at the interface of two different
materials on total internal reflection. Generally, its magnitude
is in order of the Fermi wavelength.

Recently, interesting results were reported on
the quantum GH shift for charge carriers in graphene
systems~\cite{JRZH14,JWZM15,CZL13,AM14}. This does not only
reflects the unique transport properties of Dirac electrons and
holes in graphene nanostructures, but also promotes the
application of graphene nanostructure in nanoelectronic devices.
Motivated by recent experiments on graphene superlattices (SLs)~\cite{ZD12,NLW.08,BVP09},
we consider a system of Dirac fermions through
a periodic potential in
graphene. 
We analyze the GH shifts
of the transmitted electron beam scattered by the potential
profile presented in Figure~\ref{system} through AA-staked BLG.
After formulating our model we compute the associated energy
eigenvalues and energy bands. We found that the energy bands are
just the double copies of single layer graphene bands shifted up
and down by the interlayer coupling $\gamma$. In addition, the
bounds structure of graphene SLs can have more than one Dirac
points, those locations do not depend on the number of barriers.
Then,
we used the transfer matrix method to determine the GH shifts and
 associated transmission probability. The numerical results
show that the manifestation of Klein tunneling occur at normal
incidence. Subsequently, we found that the GH shifts display sharp
peaks inside the transmission gap around the two Dirac points
($E=V_B+\tau$, $E=V_W+\tau$), where the number of this peaks is
equal to that of transmission resonances. Our findings are compared to those
of single, double barrier
structures and graphene SLs with a defect.

This paper is organized as follows. 
In section 2, we consider Dirac fermions in AA-stacked BLG system
scattered by barrier potential (Figure~\ref{system}).
In section 3, we obtain the spinor solution corresponding to each
regions composing our system. We use the transfer matrix at
boundaries together with the incident, transmitted and reflected
currents to end up with two transmission probabilities. In section
4, we numerically present
our results for the GH shifts and the transmission probability of
an electron beam transmitted through graphene SLs.  Comparison with
other graphene systems will done and  the influence of the defect mode on
our graphene SLs will be analyzed. We
conclude our work and emphasize our main results in final section.

\section{Theoretical model}

\begin{figure}[h!]
 \centering
 \includegraphics[width=11.5cm, height=4cm]{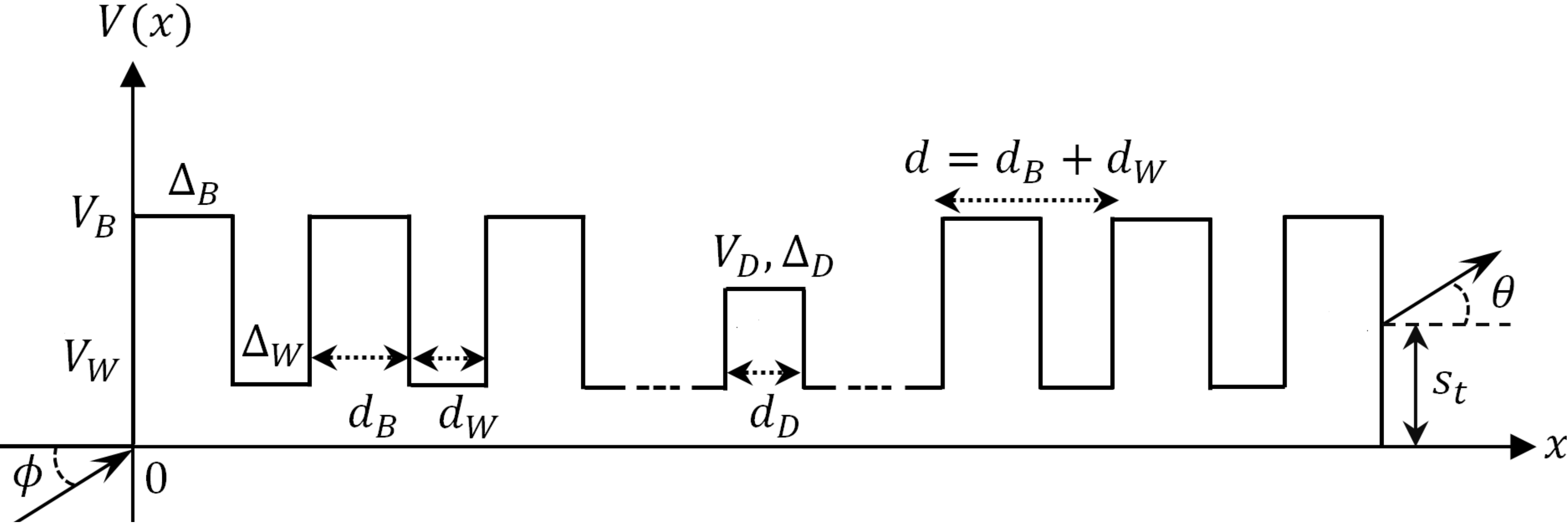}
 \caption{\sf{Schematic representation of the potential profile.
 }}\label{system}
\end{figure}

We consider an
AA-stacked BLG graphene SLs
(graphene under periodic potential) with rectangular barriers
grown along the \textit{x}-direction.
The potential profile is shown in Figure~\ref{system}, where the
symbols $"B"$, $"W"$ and $"D"$ denotes the barrier, well and the defect,
respectively. The structure are characterized by the potential
barrier height $V_B$ with gaps $\Delta_B$ and width $d_B$, the
potential well height $V_W$ with gaps $\Delta_W$ and width $d_W$.
The defect is denoted by the potential height $V_D$ with gaps
$\Delta_D$ and width $d_D$. The incidence and transmission regions
correspond to the gapless graphene with $\Delta=V=0$. Note that,
similar potential profile have been recently considered in
monolayer graphene in \cite{CZL13}.

In the basis of $\psi=\left(\psi_{A},\ \psi_{B},\ \psi_{A^{'}},\
\psi_{B^{'}}\right)^T$, where $\psi_{A(A^{'})}$ and $\psi_{B
(B^{'})}$ are the envelope functions associated with the
probability amplitudes of the wave functions on the A(A$^{'}$)
and B(B$^{'}$) sublattices of the upper (lower) layer,
our system can be described by a single valley Hamiltonian
\beq \lb{H1}
\mathcal{H}=\left(%
\begin{array}{cccc}
  V(x)+\Delta(x) & v_F(p_x-ip_y) & \gamma & 0  \\
  v_F(p_x+ip_y) &  V(x)-\Delta(x) & 0 & \gamma \\
  \gamma &  0 & V(x)+\Delta(x) &  v_F(p_x-ip_y) \\
   0 & \gamma &  v_F(p_x+ip_y) &  V(x)-\Delta(x)  \\
\end{array}%
\right) \eeq where $p=(p_x,p_y)$ being the two-dimensional
momentum operator, the Fermi velocity $v_F=10^6 \ m/s$, the
interlayer coupling $\gamma\approx0.2\ eV$~\cite{Tabert12},
the gaps $\Delta(x)$ and
potential $V(x)$. For graphene SLs with defect
the potential is defined by
\begin{equation} \label{v} V(x) =  \left
\{ \begin{array}{ccc} V_B, &&  \;  \qquad n d \leq x \leq n d +d_B
\;
\\ V_W, &&  \;  \qquad n d + d_B\leq x \leq (n+1) d \;
\\ V_D, &&  \;  \qquad N d \leq x \leq N d+d_D
 \;
\end{array} \right.
\end{equation}
and the gap reads as
\begin{equation} \label{v} \Delta(x) =  \left
\{ \begin{array}{ccc} \Delta_B, &&  \;  \qquad n d \leq x \leq n d
+d_B \;
\\ 0, &&  \; \text{otherwise}  \;
\end{array} \right.
\end{equation}
where $0\leq n \leq N-1$, $N$ being the number of period, such
that the period is considered as the alternating barriers and
wells with the width $d$ ($d= d_B+d_W$).

Due to the translation invariance in the \textit{y}-direction, the
momentum
is a conserved quantity, then the eigenvalue 
of
\eqref{H1} takes the form
\begin{equation}\label{eq2}
\psi(x,y)=e^{ik_yy}\psi(x,k_y).
\end{equation}
Solving
the eigenvalue equation $\mathcal{H} \psi = E \psi$, one finds the
energy bands
\begin{equation}\label{E-bands}
\epsilon^{\textcolor[rgb]{1.00,0.00,0.00}{s_i},\tau}_i=\tau+s_{i}\sqrt{\eta^2\left((k_{x_i}^{\tau})^2+(k_{y})^2\right)+\Delta_{i}^2}
\end{equation}
where $\epsilon_i=E-V_i$, the index $i\ (i=B,\ W,\ D)$ corresponds
to the barrier, well and defect regions,
\beq
k_{x_i}^{\tau}=\sqrt{-k_{y}^2+\eta^{-2}
\left(\left(\epsilon_i-\tau\right)^2-\Delta_{i}^2\right)}
\eeq
is
the wave vector along the \textit{x}-direction with $\tau$ is the
cone index such that $\tau=+1$ ($\tau=-1$) for the upper (lower)
cone and $s_i= \mbox{sign}(\epsilon_i)$. We further introduce the
length scale $\eta=\frac{\hbar v_F}{\gamma}\thickapprox 3.29~nm$
and switch to dimensionless quantities by measuring all energies
terms in units of the interlayer coupling $\gamma$ such that
$\epsilon_i \longrightarrow \frac{\epsilon_i}{\gamma}$,
$\Delta_i\longrightarrow \frac{\Delta_i}{\gamma}$. For the incidence
and transmission regions where we have $V = \Delta=0$, the energy
bands are
\begin{equation}\label{energybands0}
E^{s_0,\tau}=
\tau+s_0\sqrt{\eta^2\left((k_{x_0}^{\tau})^2+(k_{y})^2\right)}\end{equation}
and the wave vector reads as \beq
k_{x_0}^{\tau}=\sqrt{-k_{y}^2+\eta^{-2}\left(E-\tau\right)^2}\eeq
with $s_0=\mbox{sign}(E).$

\begin{figure}[h!]
 \centering
 \includegraphics[width=5.8cm, height=4.5cm]{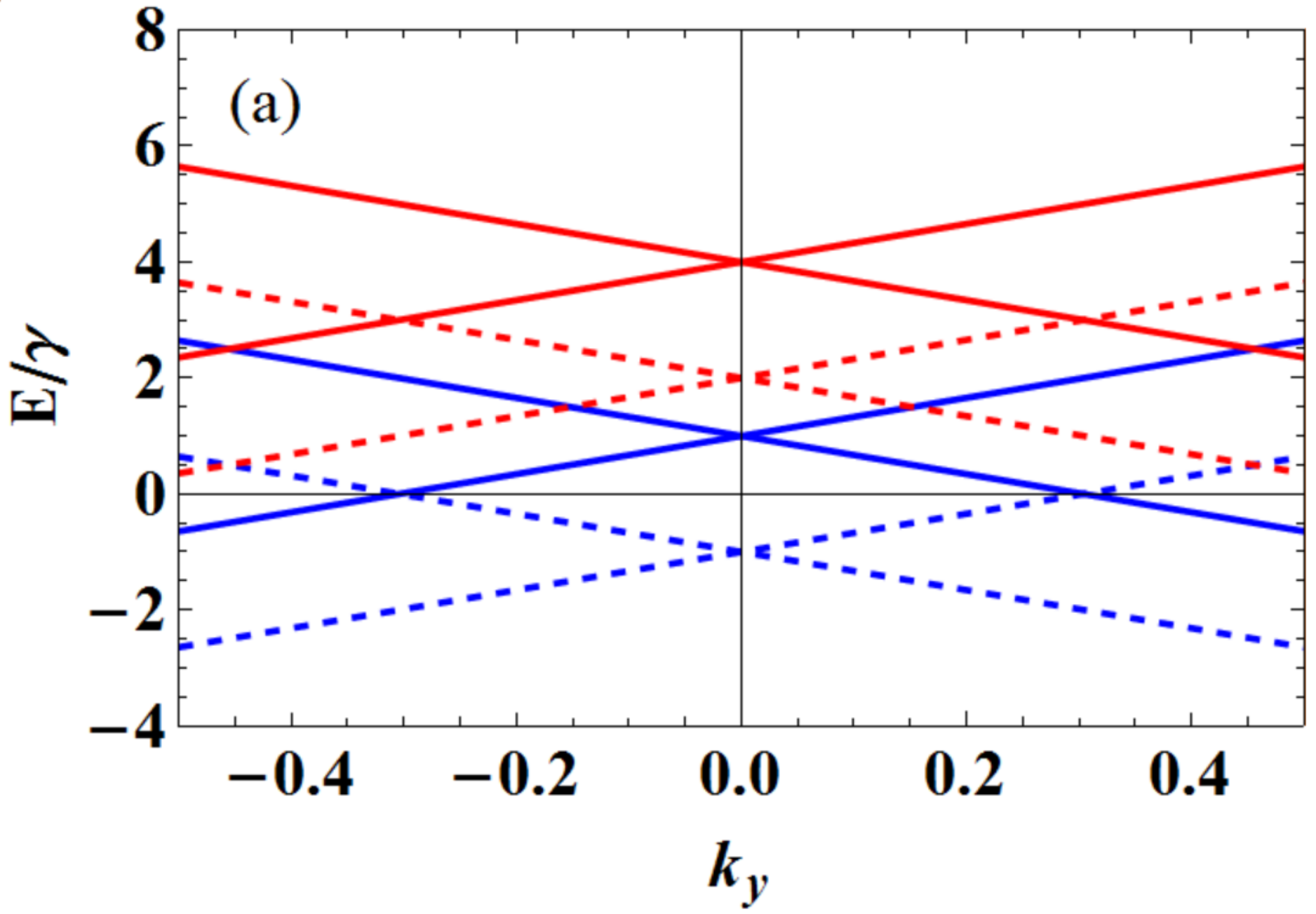}
 \qquad \ \ \ \includegraphics[width=5.8cm, height=4.5cm]{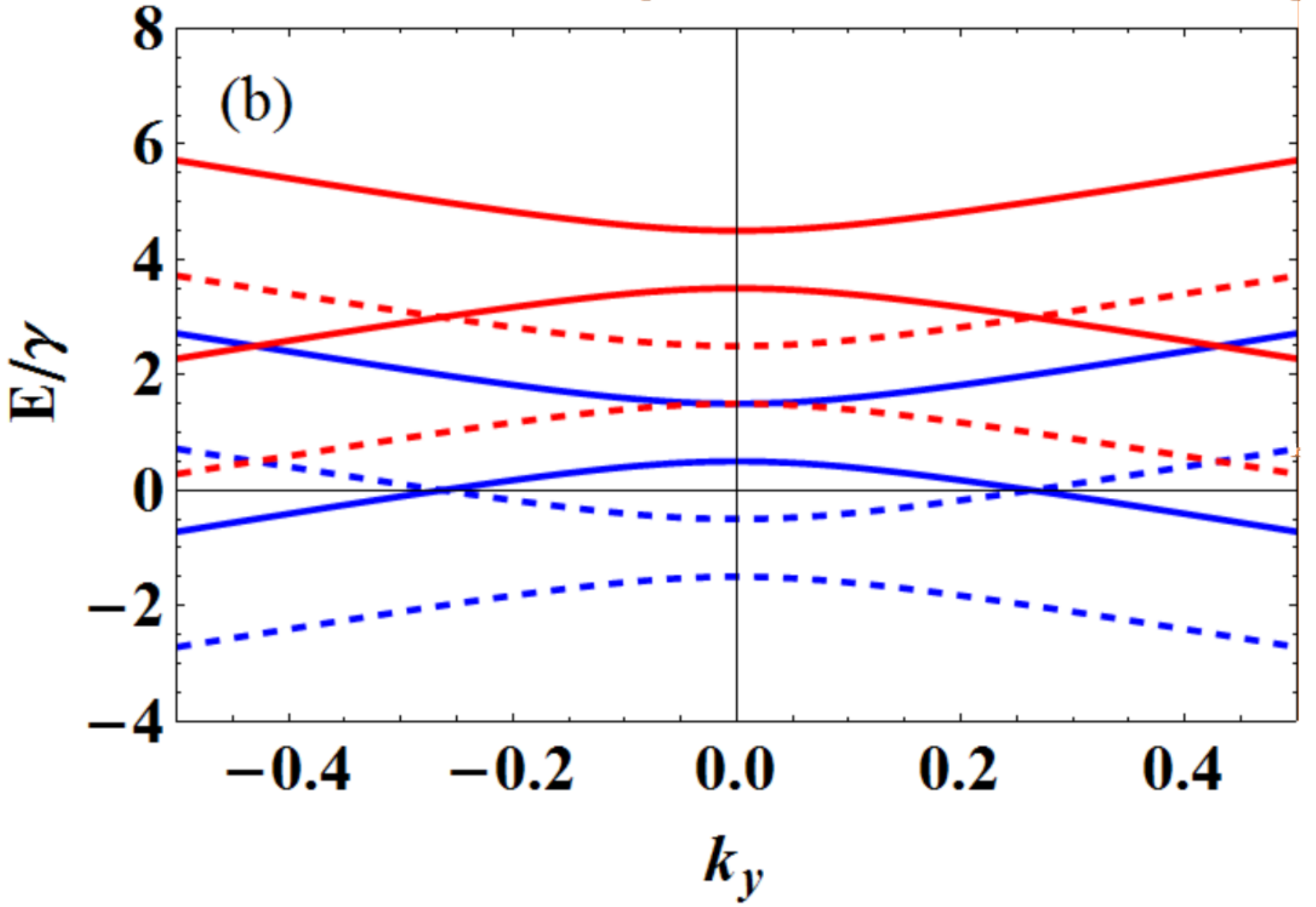}
 \caption{\sf{Energy bands as function of the momentum $k_y$. The solid and  dashed curves
correspond to the upper and  lower cone \textit{i.e.}
$\tau=+1$ and $\tau =-1$, respectively, where the physical
parameters are chosen to be, (a) : $V_B=0$ (blue curves), $V_B=3
\gamma$ (red curves) and $\Delta_B=0$. (b) : $V_B=0$ (blue
curves), $V_B=3 \gamma$ (red curves) and $\Delta_B=0.5
\gamma$.}}\label{energy}
\end{figure}

Plots of the energy bands structure in AA-stacked BLG,
evaluating~\eqref{E-bands} for the barrier region, have been shown
in Figure~\ref{energy}. We clearly see that the energy bands are
different from that of the AB-stacked BLG~\cite{VP13} and also
monolayer graphene~\cite{NGMJ.04}.
One can observe that for zero gap, the energy bands are linear and
just two copies of the monolayer band structure shifted up and
down by $\gamma=0.2 \ eV$ (Figure~\ref{energy}(a)), respectively.
In addition,
the Dirac points
are located at $E=V_B \pm
\ \gamma$.
For a finite gap, the spectrum is parabolic and the two Dirac
points are lifted and they are shifted up and down by $\Delta_B$
(Figure~\ref{energy}(b)).
We can observe that when
the potential heights $V_B$ increase, the energy bands increase
upwards.


As usual, to derive the eingespinors we solve $\mathcal{H} \psi=
E\psi$. The corresponding eigenspinors can be written as
\begin{equation}
\psi_{i}=L_i \cdot A_i
\end{equation}
where we have set
\begin{equation}\lb{solM}
L_{i}=\left(%
\begin{array}{cccc}
s_i f_{i}^{+,+} e^{ik_{x_i}^{+}x}&  s_i f_{i}^{+,-}
e^{-ik_{x_i}^{+}x}& s_i f_{i}^{-,+}e^{ik_{x_i}^{-}x}  & s_i
f_{i}^{-,-} e^{-ik_{x_i}^{-}x} \\ e^{ik_{x_i}^{+}x} & e^{-ik_{x_i}^{+}x} & e^{ik_{x_i}^{-}x} & e^{-ik_{x_i}^{-}x} \\
s_i f_{i}^{+,+}e^{ik_{x_i}^{+}x} &  s_i f_{i}^{+,-}
e^{-ik_{x_i}^{+}x}& -s_i f_{i}^{-,+}e^{ik_{x_i}^{-}x}  & -s_i
f_{i}^{-,-} e^{-ik_{x_i}^{-}x}\\ e^{ik_{x_i}^{+}x}  &
e^{-ik_{x_i}^{+}x} &  -e^{ik_{x_i}^{-}x} & -e^{-ik_{x_i}^{-}x}
\end{array}%
\right),
\qquad A_{i}=\left(%
\begin{array}{cc}
  \alpha_{i}^+ \\
  \beta_{i}^+  \\ 
    \alpha_{i}^- \\
  \beta_{i}^-
  \end{array}%
\right)
\end{equation}
with
\beq
f_{i}^{\tau,\pm}=\pm\sqrt{\frac{E-V_i-\tau+\Delta_i}{E-V_i-\tau-\Delta_i}}e^{\mp
i\phi_{i}^\tau}, \qquad \phi_{i}^\tau=\arctan(k_y/k_{x_i}^\tau).
\eeq
It may be noted that for AB-stacked BLG, in the four band model,
we have four reflection and four transmission
channels~\cite{VP13}.
For AA-stacked BLG, all intercone transitions ($\tau
\longrightarrow - \tau$) are strictly forbidden due to the
orthogonality of electron wave functions with a different cone
index~\cite{SAZ13}, which yield to only two transmissions ($\tau
\longrightarrow \tau$). Thus, we can reduce the $4\times4$ matrix
to the following $2\times2$ matrix
\begin{equation}\lb{L}
L_{i}=\left(%
\begin{array}{cc}
s_i\tau f_{i}^{\tau,+} e^{ik_{x_i}^{\tau}x} &  s_i\tau
f_{i}^{\tau,-} e^{-ik_{x_i}^{\tau}x}\\ \tau e^{ik_{x_i}^{\tau}x} &
\tau e^{-ik_{x_i}^{\tau}x}
\end{array}%
\right),
\qquad A_{i}^{\tau}=\left(%
\begin{array}{cc}
  \alpha_{i}^\tau \\
  \beta_{i}^\tau  \\
  \end{array}%
\right).
\end{equation}

\section{Transmission and Goos-H\"{a}nchen shifts}


We are interested in the normalization coefficients, the
components of $A_{i}^{\tau}$, on the both sides of the
superlattices. In other words,
for the incidence and transmission regions,  we have, respectively
\beq
A_{in}^{\tau} =\left(%
1,r^{\tau} \right)^T, \qquad A^{\tau}_{tr} =\left(%
t^{\tau} ,0\right)^T
\eeq
where $r^\tau $ and $t^\tau $ are the
reflection and transmission coefficients of each cone ($\tau=\pm
1$), respectively.
The coefficients $r^\tau$ and $t^\tau$ are determined by imposing
the continuity of the wave functions. For this, we need to match
the wave functions at the boundaries between different regions.
This procedure is most conveniently expressed in the transfer
matrix formalism~\cite{Wang10}.
From the continuity of the wave function, we end up with
\beq\lb{rt}
   \left(
   \begin{array}{c}
              {1} \\
              {r^{\tau}} \\
            \end{array}
          \right)= M \left(
            \begin{array}{c}
              {t^{\tau}} \\
              {0} \\
            \end{array}
          \right)
\eeq where $M$ is the transfer matrix \beq \lb{transferM} M=
L_{in}^{-1}\left[0\right]\cdot G \cdot L_{tr}\left[ \zeta \right]
\eeq
Note that $L_{in}=L_{tr}$ is determined from~\eqref{L} at the
condition $V=\Delta=0$. The matrix $G$ and width $\zeta$ depend on
the choice of the graphene structures. For graphene SLs
($(BW)^N B (WB)^N$),  $BW$ means a barrier followed by a well
(i.e., period) and $N$ is the number of period,
$G$ is given by
 \beq \lb{G} G=(F_B\cdot F_W)^N\cdot F_B \cdot( F_W\cdot
F_B)^N
\eeq
and  $\zeta$ reads as
\beq \lb{eta} \zeta=2N d + d_B\eeq
where the matrix $F_i$ takes the form
\beq
F_{i}=\frac{1}{\cos{\phi_i}}\left(%
\begin{array}{cc}
\cos{\left(k_{x_i} d_i +\phi_i\right)} &  -i s_i f_{i}^{\tau} \sin{\left(k_{x_i} d_i  \right)}\\
-i \frac{s_i}{f_{i}^{\tau}} \sin{\left(k_{x_i} d_i \right)} &
\cos{\left(k_{x_i} d_i -\phi_i\right)}
\end{array}%
\right), \qquad i=B, W, D.
\eeq
We can explicitly write the above relations for three graphene systems. Indeed,
single barrier: 
\beq
G=F_B,\qquad \zeta=d_B
\eeq
Double barrier:
\beq
G=F_B\cdot F_W \cdot F_B, \qquad \zeta=d + d_B
\eeq
Graphene SLs ($(BW)^N D (WB)^N$) with defect $D$:
\beq
G=(F_B\cdot F_W)^N\cdot F_D \cdot( F_W\cdot F_B)^N, \qquad \zeta=2N
d + d_D.
\eeq

From the above analysis, now one can
obatin two channels for the transmission
probability in each individual cone.
These are given by
\beq
t^\tau=\frac{1}{M_{11}}
\eeq
where $M_{11}$ \textcolor[rgb]{1.00,0.00,0.00}{is} an elements of the transfer matrix $M$ given by~\eqref{transferM}. 
After a lengthy but straightforward algebra, we  show that the
transmission coefficient can be written in terms of the phase
shift $\varphi$ as
\beq \lb{t} t^\tau=\frac{1}{f_0 e^{-i \varphi}}.\eeq
Using \eqref{transferM} to end up with \beq \lb{phi}f_0 e^{i
\varphi}= \frac{1}{2}\left(G_{11}+G_{22} + i\left[
s_0\left(G'_{12}+G'_{21}\right)\sec{\phi_0}+\left(G_{22}-G_{11}\right)\tan{\phi_0}
\right]\right) \eeq where $G_{\mu \nu}$ are the matrix element of
$G$ with $G'_{\mu \nu }=-i G_{\mu \nu }$ for $\mu \neq \nu$. The
phase shift can be expressed explicitly as
\beq
\lb{phase}\varphi=\arctan\left[\frac{s_0\left(G^{'}_{12}+G^{'}_{21}\right)\sec{\phi_0}+\left(G_{22}-G_{11}\right)\tan{\phi_0}}{G_{11}+G_{22}}
\right].
\eeq

The phase obtained  obove can be used to
investigate the GH shifts for the transmitted electron beam
through the AA-stacked BLG superlattices.
Indeed we look at the GH shifts by considering an incident,
reflected and transmitted beams around some transverse wave vector
$k_y=k_{y_0}$ corresponding to the central incidence angle
$\phi=\phi_0$, denoted by the subscript 0. These can be expressed
in integral forms as
\begin{eqnarray}
   \Psi_{in}^{\tau}(x,y) &=& \int_{-\infty}^{+\infty}dk_y\ g(k_y-k_{y_0})\ e^{i(k_{\textcolor[rgb]{1.00,0.00,0.00}{x_0}}^{\tau} x+k_y y)}\left(
            \begin{array}{c}
              {s_0 \tau e^{-i \textcolor[rgb]{1.00,0.00,0.00}{\phi_{0}^{\tau}}}} \\
              {\tau} \\
            \end{array}
          \right)\label{inci}\\
\Psi_{re}^{\tau}(x,y) &=& \int_{-\infty}^{+\infty}dk_y\ r^{\tau}\
g(k_y-k_{y_0})\
e^{i(-k_{\textcolor[rgb]{1.00,0.00,0.00}{x_0}}^{\tau} x+k_y
y)}\left(
            \begin{array}{c}
              {-s_0 \tau e^{-i \textcolor[rgb]{1.00,0.00,0.00}{\phi_{0}^{\tau}}}} \\
              {\tau} \\
            \end{array}
          \right)\label{refl}\\
   \Psi_{tr}^{\tau}(x,y) &=& \int_{-\infty}^{+\infty}dk_y\ t^{\tau}\ g(k_y-k_{y_0})\ e^{i(k_{\textcolor[rgb]{1.00,0.00,0.00}{x_0}}^{\tau} x+k_yy)}\left(
            \begin{array}{c}
              {s_0 \tau e^{-i \textcolor[rgb]{1.00,0.00,0.00}{\phi_{0}^{\tau}}}} \\
              {\tau} \\
            \end{array}
          \right)\label{trans}
\end{eqnarray}
where each spinor plane wave is
associated to the Hamiltonian~\eqref{H1} and $g(k_y-k_{y_0})$ is
the angular spectral distribution, which  assumed to be of
Gaussian shape.
To calculate the GH shifts of the transmitted beam through our
system,
according to the stationary phase method~\cite{B51}, we adopt the
definition \cite{JRZH14,AM14}
\beq \lb{st}
s_t^{\tau}=-\frac{\partial \varphi (k_y)}{\partial k_{y_0}}
\eeq
 where the phase $\varphi (k_y)$ of the transmission
coefficient $t^{\tau}(k_y)$, defined in \eqref{phase}, depend on
$k_y$. Note that, we have two shifts corresponding to the upper
($\tau=+1$) and lower ($\tau = -1$) cones.

\section{Numerical results}

We will numerically study
the transmission and GH shifts
in different graphene based nanostructures: single, double
barriers and graphene SLs with and without defect.
In what follow, the GH shifts well be calculated in unit of the
Fermi wavelength with  the length scale $\eta$
\beq \lambda=\frac{2 \pi\eta }
{E-\tau}.
\eeq


\subsection{Perfect transmission at normal incidence}

\begin{figure}[h!]
 \centering
 \includegraphics[width=5.5cm, height=4.5cm]{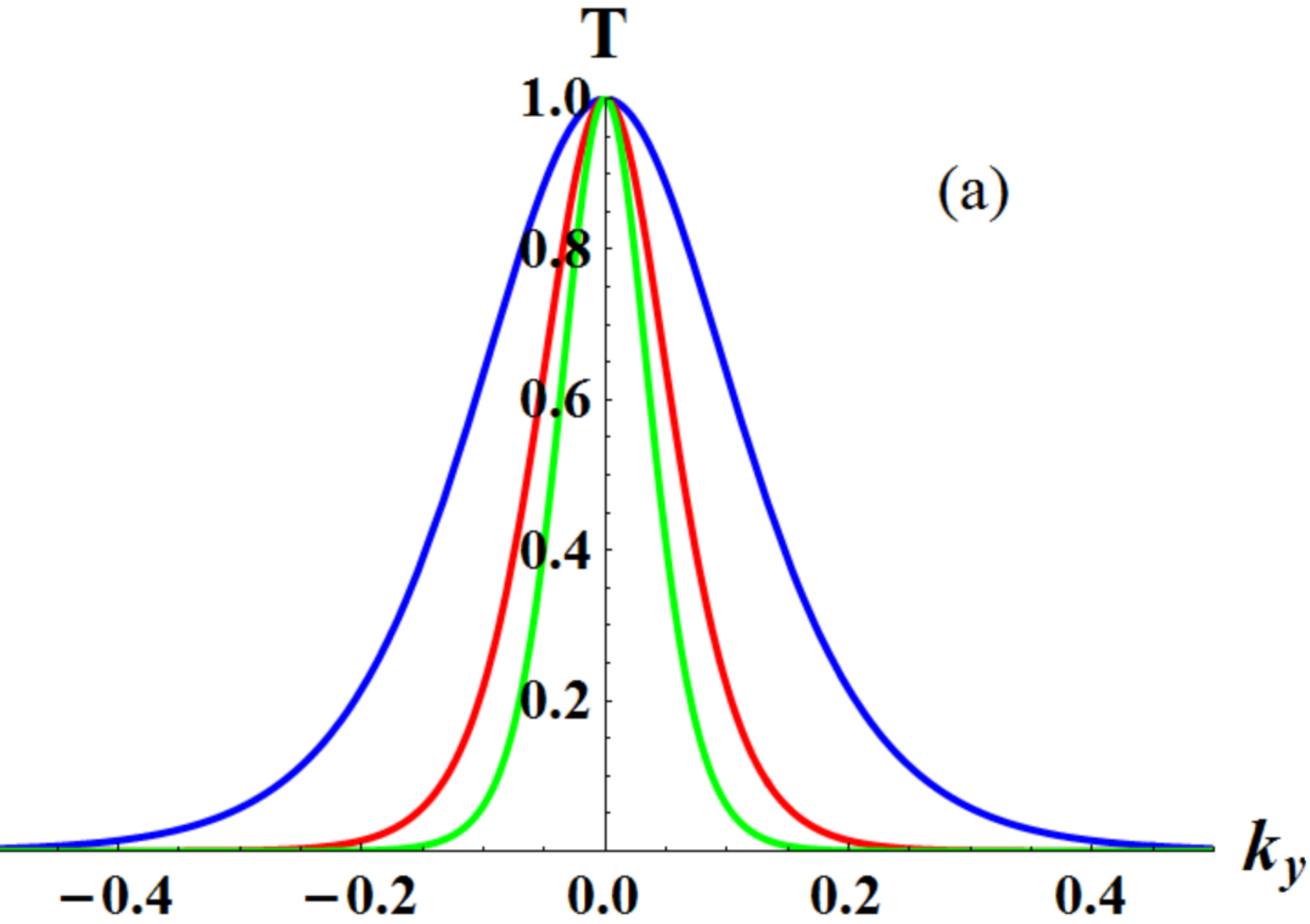}
 \quad \ \ \includegraphics[width=5.5cm, height=4cm]{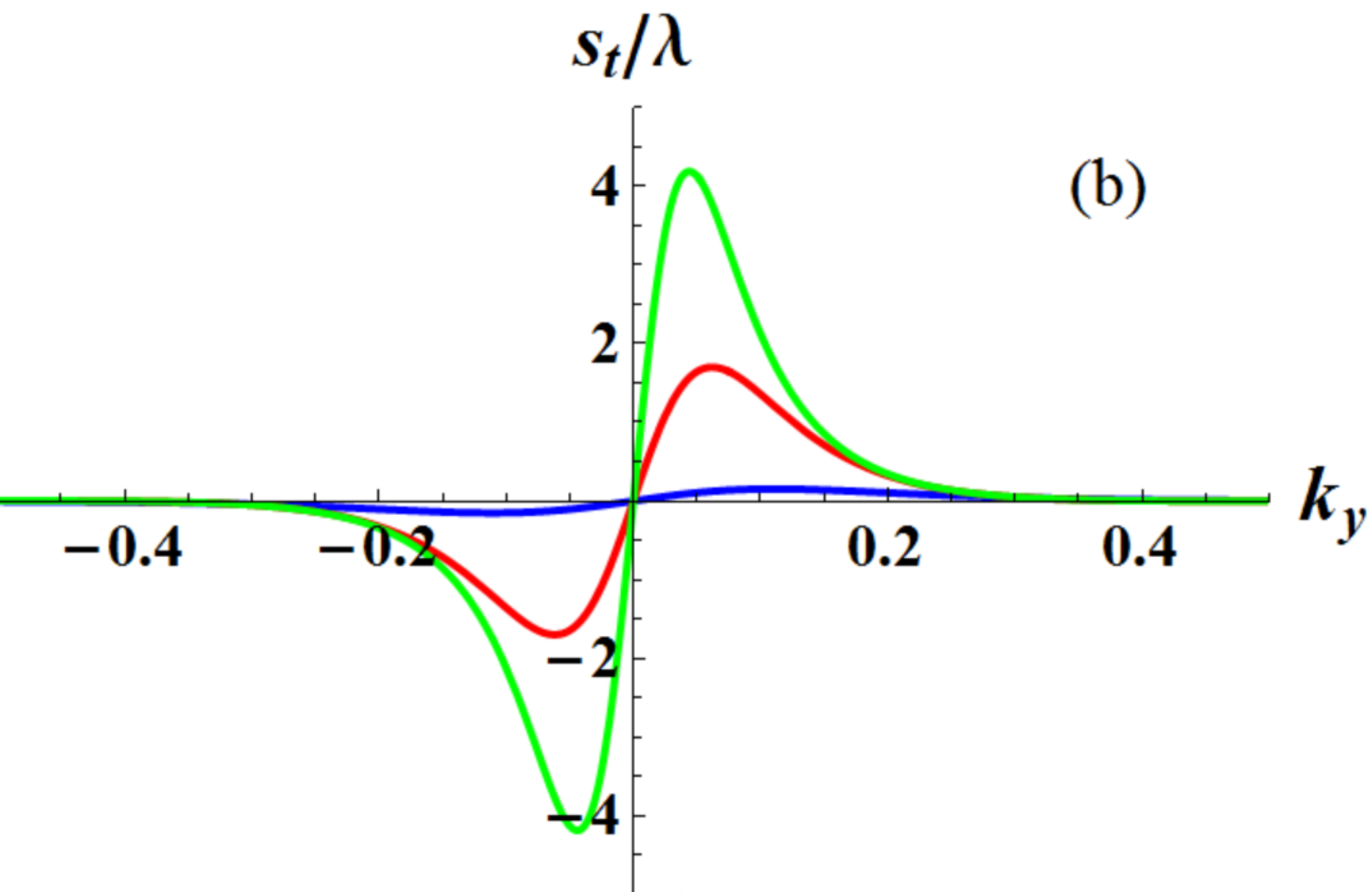}\\
\caption{\sf{Transmission probability (a) and Goos-H\"{a}nchen
shifts (b) as function of transverse wave vector $k_y$, around the
Dirac point ($E=V_B + \tau$), for
 single (blue line), double (red line) barrier and graphene superlattices (green line), where physical parameters are chosen to be
 $\Delta_B=0$, $d_W=  d_B =7 nm$, $V_B=10 \gamma$ and $V_W=4 \gamma$.}}\label{TGHS-ky}
\end{figure}

We recall that for
monolayer graphene,
the Klein tunneling
(perfect transmission) manifest at normal incidence through
potential barriers as predicted in~\cite{KNG06} and experimentally
observed~\cite{YK09,SHG09}. While, in the case of AB-stacked BLG
there is no Klein tunneling
at normal incidence~\cite{KNG06,BVP10}. Now we turn our attention
to investigate the basic behaviors of
single, double barriers and graphene SLs of AA-stacked BLG for
zero gap. Figure~\ref{TGHS-ky}, presents the transmission and GH
shifts as a function of the wave vector $k_y$ around the Dirac
point ($E=V_B + \tau$) with
$\Delta_B=0$, $d_W=  d_B =7 nm$, $V_B=10 \gamma$ and $V_W=4
\gamma$. The blue, red and green lines correspond to single,
double barriers and graphene SLs composed of 7 regions,
respectively.
\begin{itemize}
 \item
From Figure~\ref{TGHS-ky}(a), we clearly see that the transmission
exhibits a maximum (perfect transmission) for normal incidence,
($k_y=0$) and vanishes for specific values that decrease by
increasing the number of barriers. We observe that the curve of
$T$ is bilaterally symmetrical with respect to the normal
incidence around the Dirac point.
Our transmission for the upper layer $T^+$ is equal to
the transmission for the lower layer $T^-$, which give the
total transmission as average.
Recall that,
for AB-stacked BLG the total transmission is
resulting from four transmission channels~\cite{VP13}.
\item
Figure~\ref{TGHS-ky}(b) shows that the GH shifts change the sign
at the Dirac point $E=V_B + \tau$.
We observe that by increasing the number of barriers, the maximum
absolute values of the GH shifts increase. In contrast to the transmission, the
shifts is bilaterally asymmetrical with respect to the normal
incidence around the Dirac point. Since the shifts are strongly
related to the transmission,
we can conclude that the total
shifts are the average of the two shifts corresponding to the
upper and lower cones.
\end{itemize}

\subsection{Single barrier structure}

\begin{figure}[H]
 \centering
\ \includegraphics[width=7cm, height=6cm]{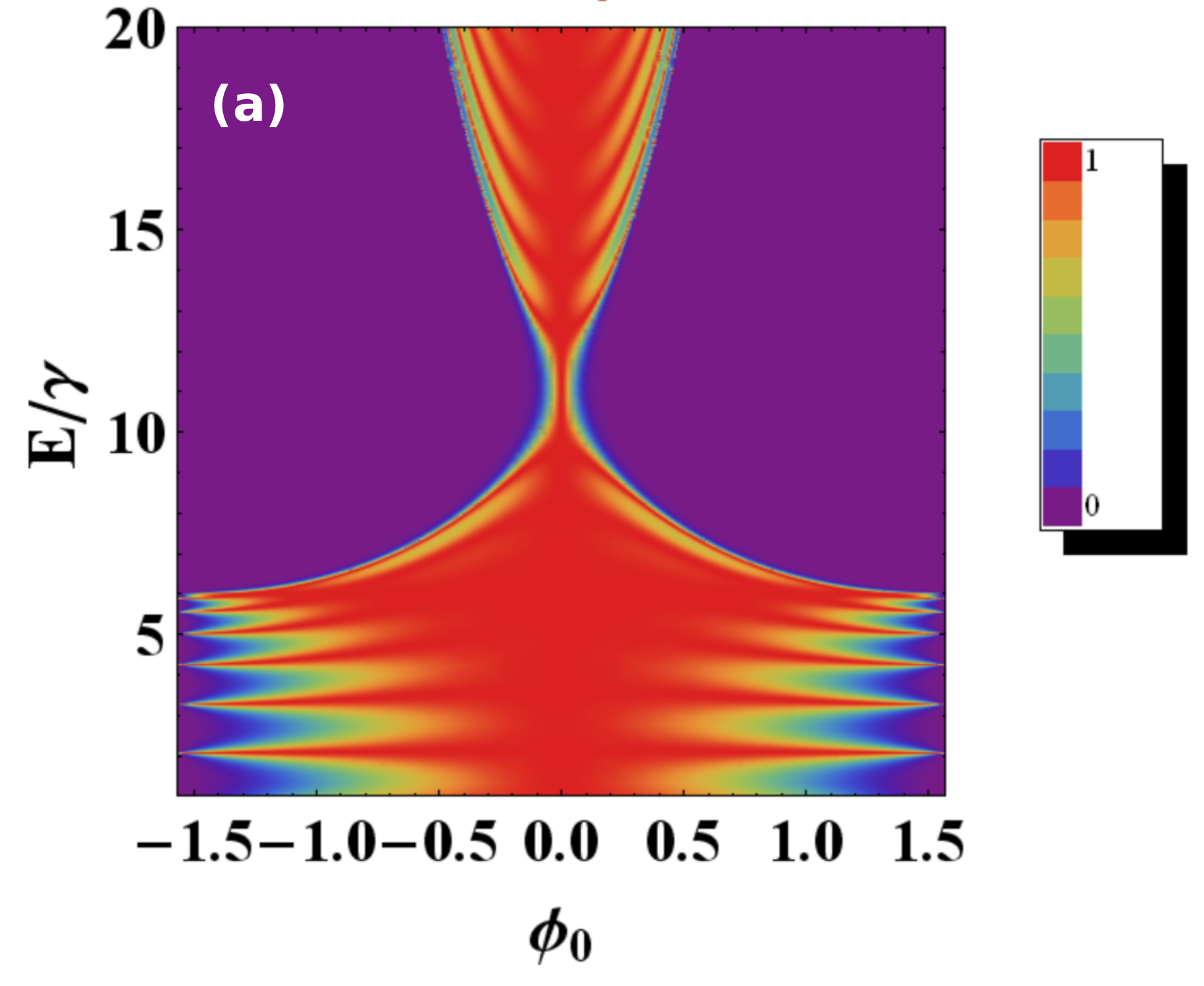}
 \quad \ \ \includegraphics[width=7cm, height=6cm]{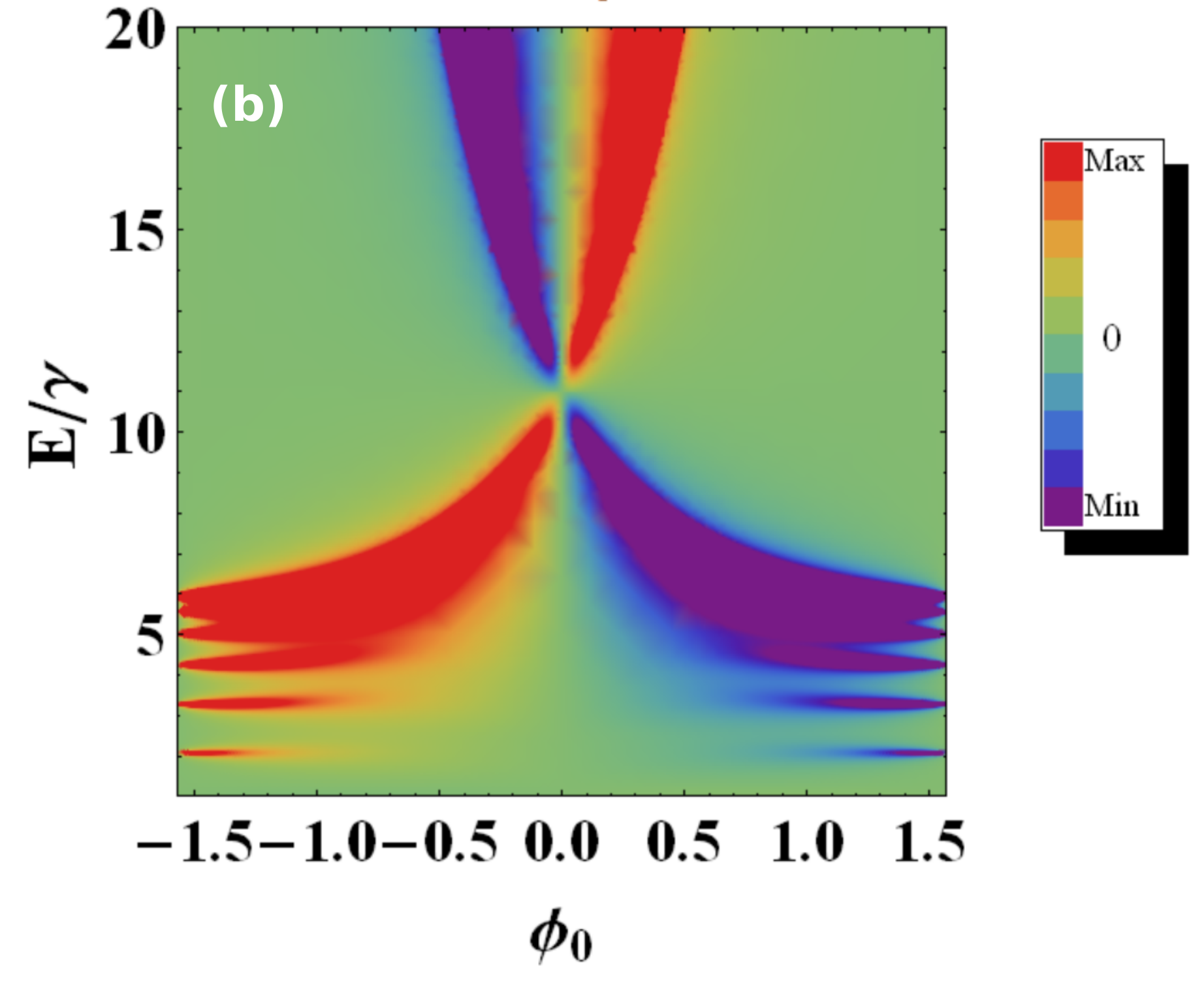}\\
 \ \ \ \includegraphics[width=7cm, height=6cm]{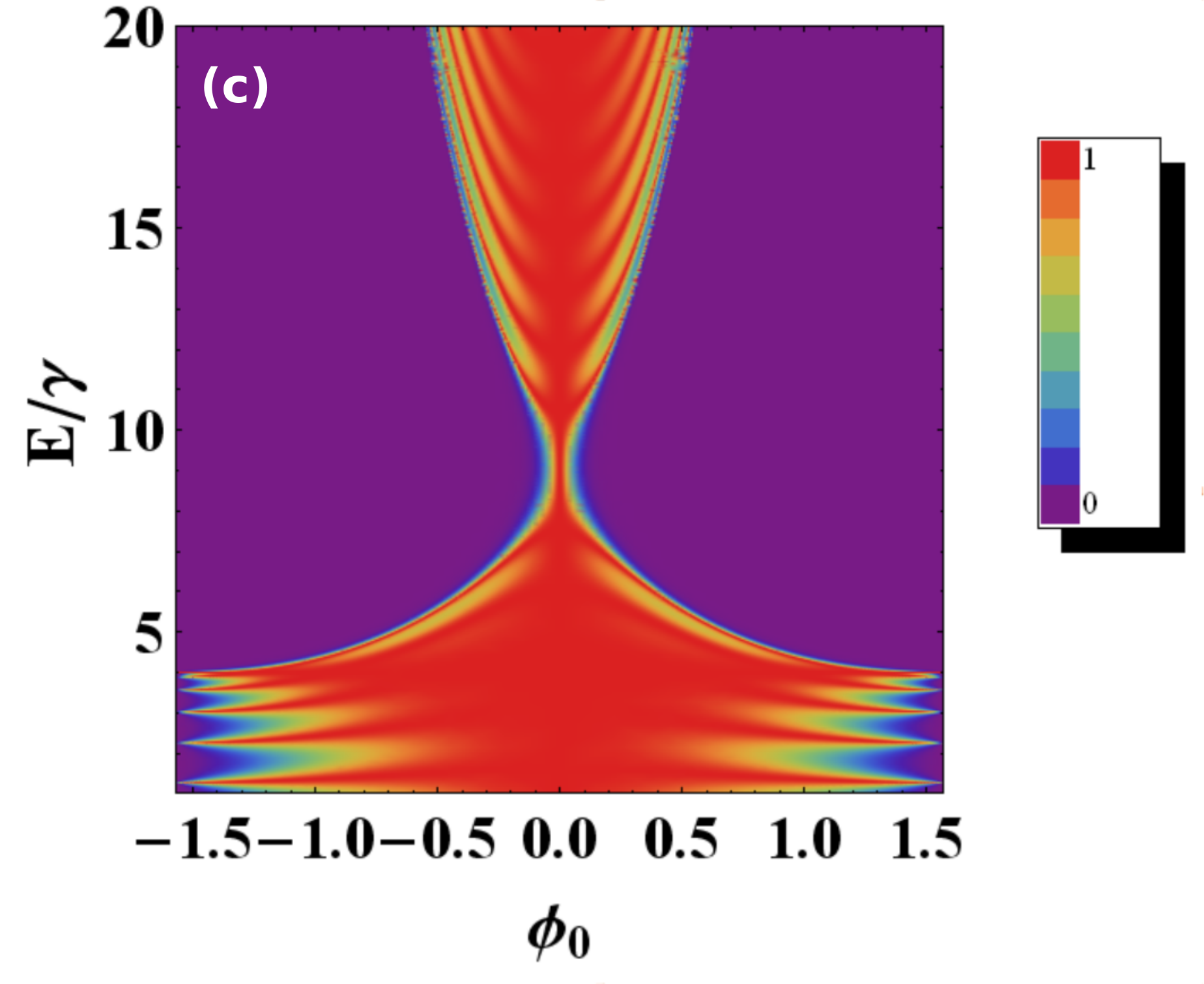}
\ \quad   \  \includegraphics[width=7cm, height=6cm]{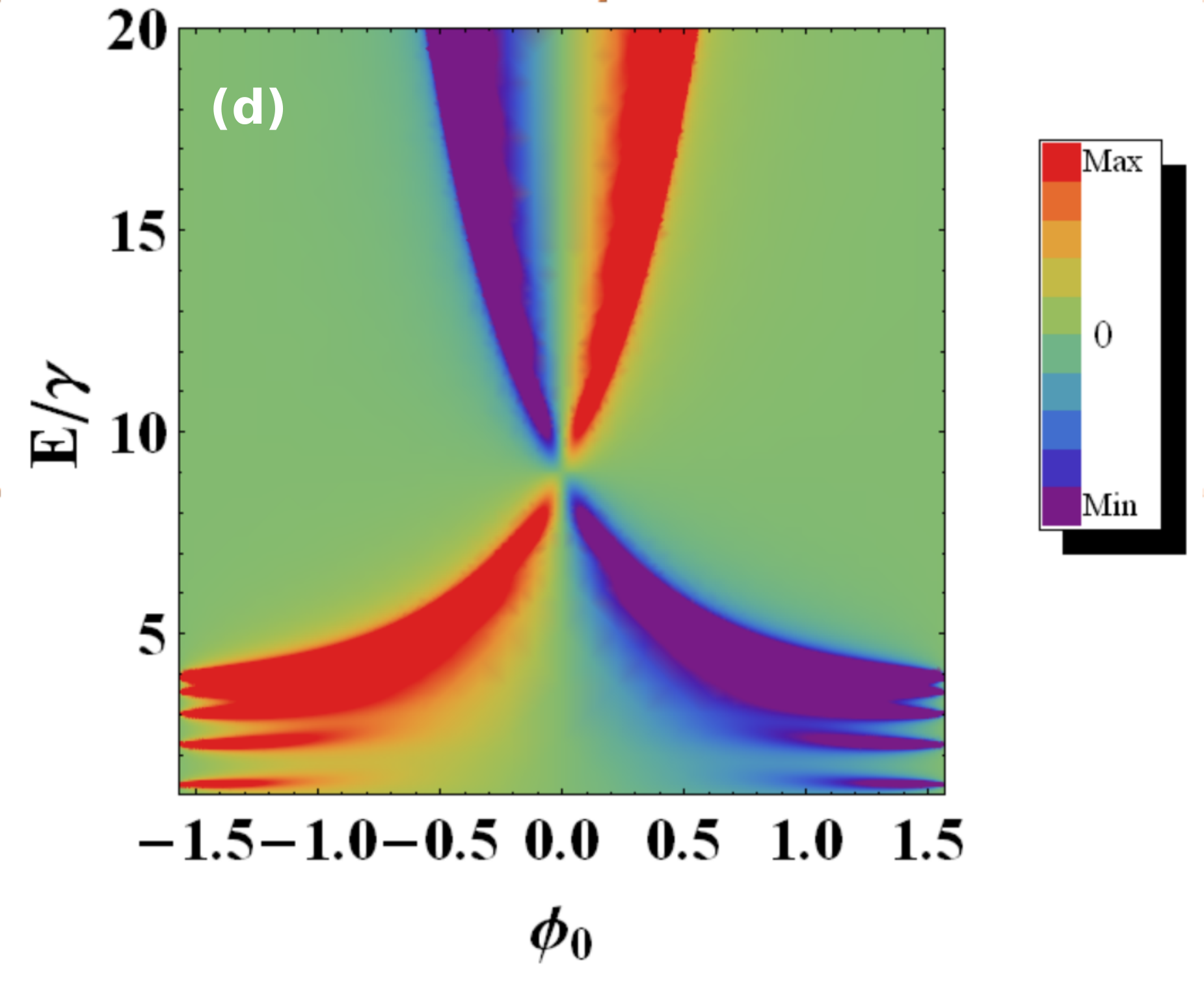}
 \caption{\sf{Density plot of the transmission probability and Goos-H\"{a}nchen shifts as function of the energy for single
 barrier, with physical parameters
 $\Delta_B=0$, $d_B=7 nm$ and $V_B=10 \gamma$. $\tau =1$ for (a, b),  $\tau =-1$ for (c, d).}}\label{TGHS-1b}
\end{figure}

We consider electron in AA-stacked BLG scattered by
single barrier structure in the absence of the gap.
In Figure~\ref{TGHS-1b}, {we
present}
the density plot of the
transmission probability
and GH shifts as a function of the incident angle and its energy
corresponding to the upper ($\tau=+1$) and lower ($\tau=-1$)
cones, for $\Delta_B=0$, $d_B=7 nm$ and $V_B=10 \gamma$. Recall
that for AA-stacked BLG the band structure is composed of two
Dirac cones shifted by $\tau$. From Figure~\ref{TGHS-1b}, we
see that the transmission and the GH shifts, for both cones, has
the same form as that in the case of monolayer
graphene~\cite{AM14}. Obviously, the GH shifts can be positive as
well as negative and are closely related to the transmission gap.
\begin{itemize}
 \item
Figure~\ref{TGHS-1b}(b) and \ref{TGHS-1b}(d) indicates
that, for positive incident angle, the GH shifts are negative for
$E< V_B + \tau$, positive for $E> V_B + \tau$, change the sign
near the Dirac point,
$E=V_B + \tau$ and
become large at some resonance points.
Since the GH shifts are bilaterally asymmetrical with respect to
the normal incidence, for negative incident angle, the shifts are
positive for $E< V_B + \tau$ and negative for $E> V_B + \tau$.
\item
Figures~\ref{TGHS-1b}(a) and \ref{TGHS-1b}(c) show that there is
perfect transmission for normal or near normal incidence
($\phi_0^{\tau} \rightarrow 0$), which is a manifestation of the
Klein tunneling~\cite{KNG06}.
We notice that the angular dependence of the transmission
probability  is very remarkable. Moreover, the
transmission is bilaterally symmetrical  with respect to the normal incidence.
\end{itemize}



\subsection{Double barrier structure}

\begin{figure}[H]
 \centering
 \includegraphics[width=7cm, height=6cm]{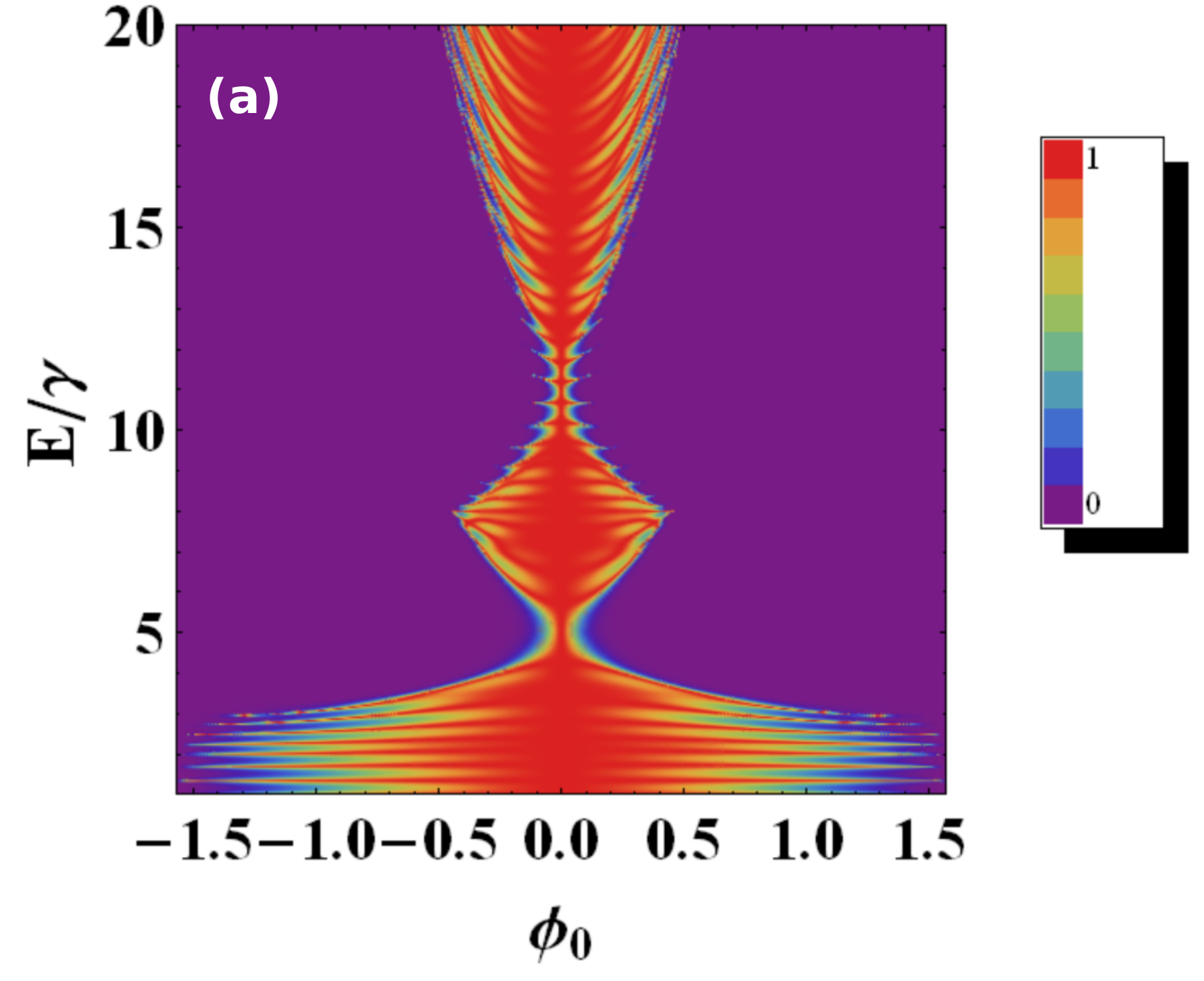}
 \quad \ \ \includegraphics[width=7cm, height=6cm]{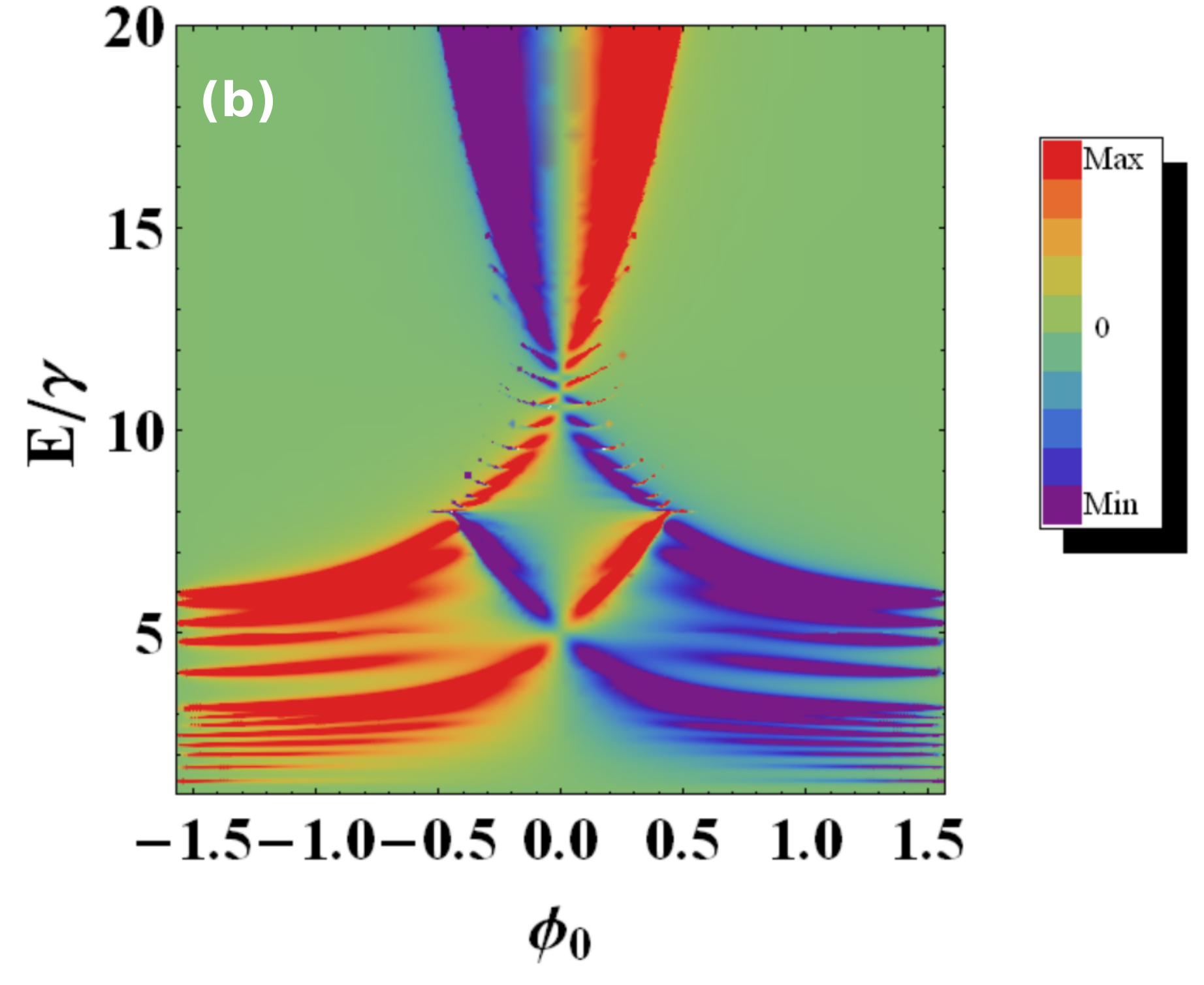}\\
   \includegraphics[width=7cm, height=6cm]{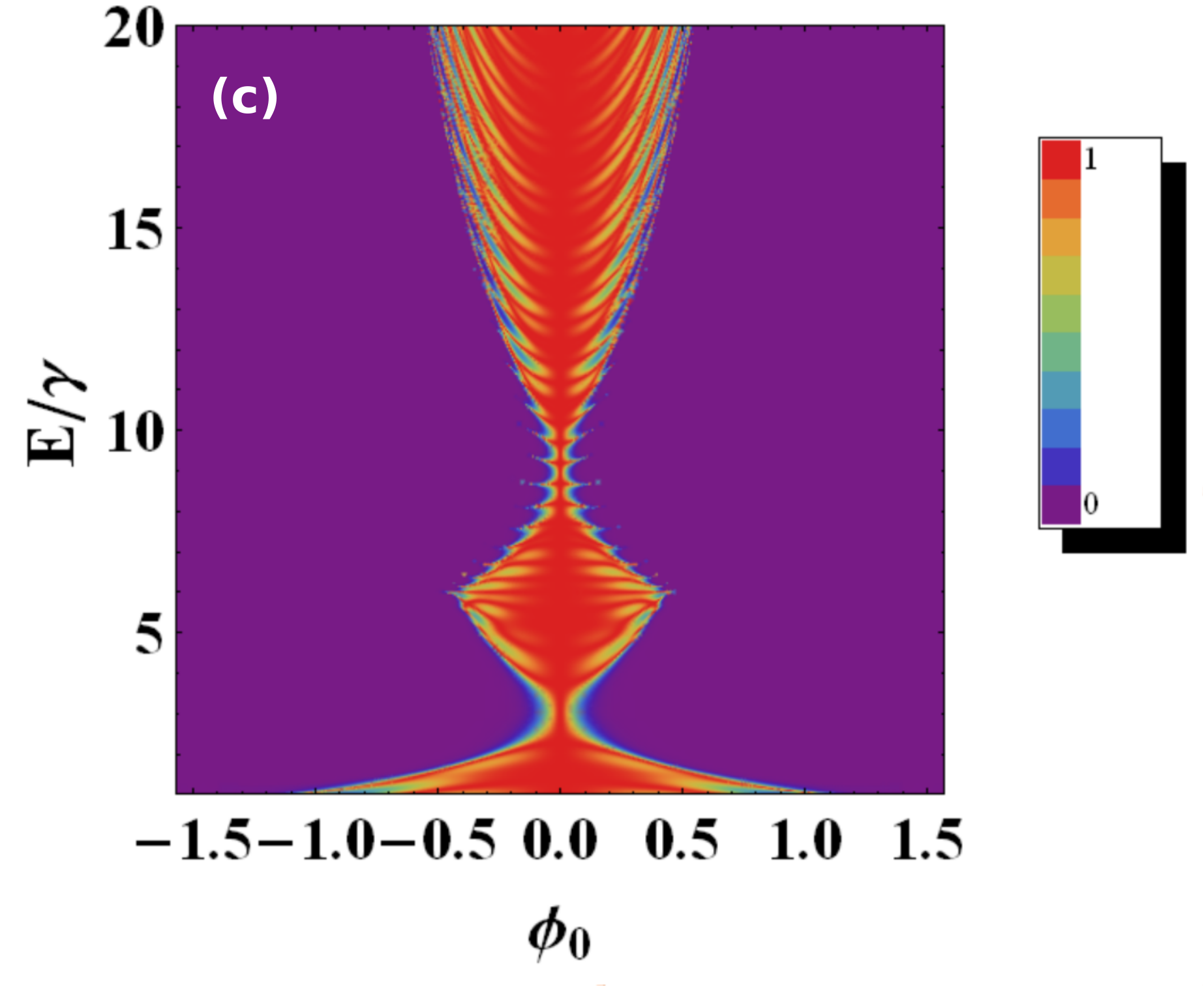}
\quad  \ \ \includegraphics[width=7cm, height=6cm]{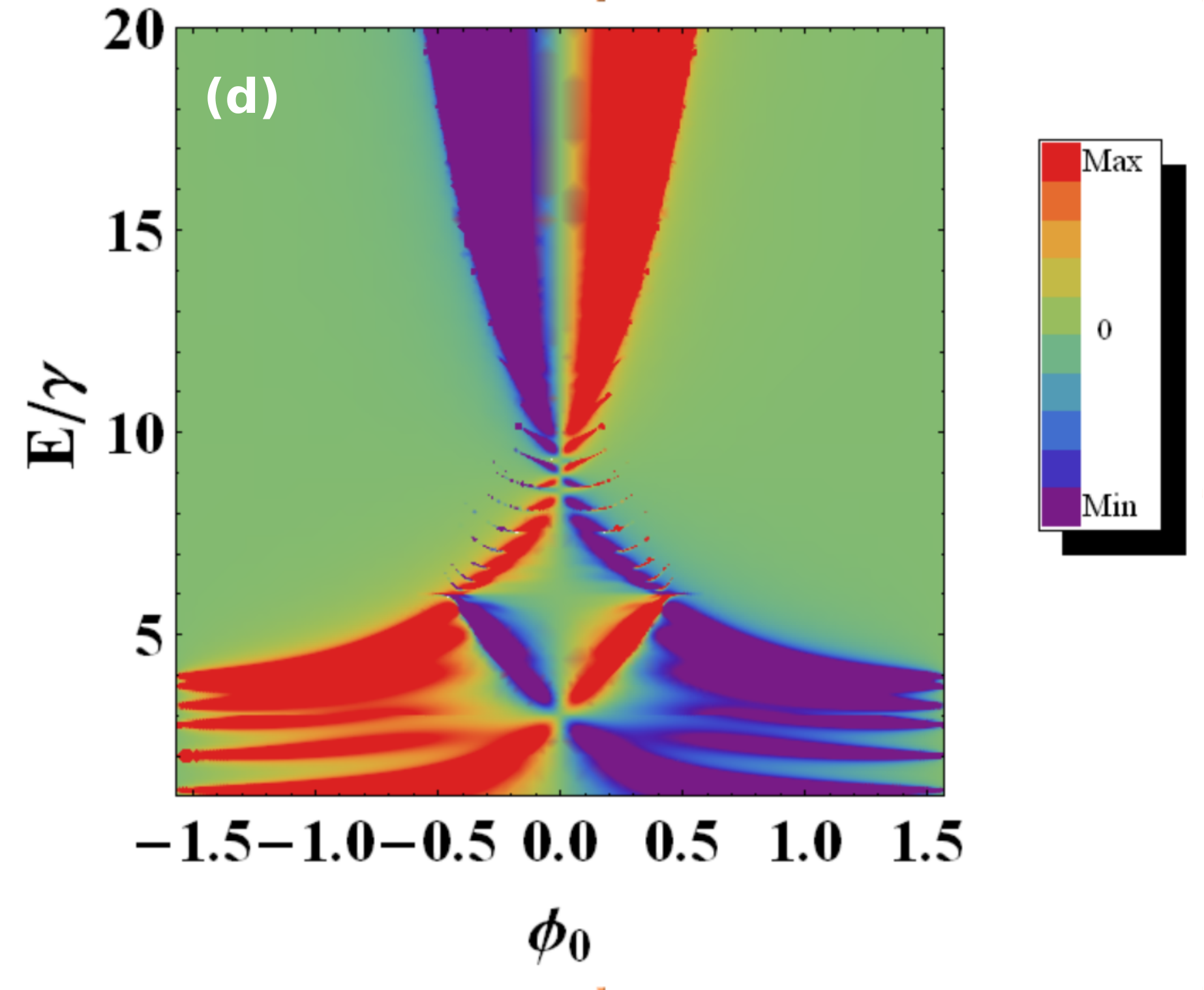}
 \caption{\sf{Density plot of the transmission probability and Goos-H\"{a}nchen shifts as function of the incident angle and its energy for
 double barrier structure, with physical parameters
 $\Delta_B=0$, $d_W= 2 d_B =14 nm$, $V_B=10 \gamma$ and $V_W=4 \gamma$.
  $\tau =1$ for (a,b),  $\tau =-1$ for (c,d).}}\label{TGHS-2b}
\end{figure}

Now we consider a double barrier structure,
the transmission probability and GH shifts as a function of the
incident angle and its energy for both cones are presented in
Figure~\ref{TGHS-2b}. The physical parameters are chosen to be
$\Delta_B=0$, $d_W= 2 d_B =14 nm$, $V_B=10 \gamma$ and $V_W=4
\gamma$.

\noindent Figures~\ref{TGHS-2b}(a,c) correspond to the
transmission, while~\ref{TGHS-2b}(b,d) correspond
to the GH shifts for the upper ($\tau=+1$) and lower ($\tau=-1$)
cones, respectively.
Compared to the results of single barrier, there is a new Dirac
point, which appears at $E=V_W + \tau$
that is resulting from the chiral nature of massless Dirac
excitations. It is clearly seen that the transmission displies
sharp peaks inside the transmission gap around the Dirac point
located at $E=V_B +\tau$, while they are absent around the new
Dirac point located at $E=V_W +\tau$ that corresponds to the first
transmission gap. We notice that, the transmission resonances are
resulting from the available states in the well between the
barriers. For positive incident angle, the GH shifts are negative
for $E< V_W + \tau$ and positive for $E> V_B +\tau$. However, for
$V_W +\tau < E < V_B + \tau$ the shifts shows different behaviors,
which are positive and negative, respectively. In addition, the
shifts display sharp peaks, that are equal to that of the
transmission resonances, inside the transmission gap around
$E={V_B} +\tau$. While, these peaks
are absent around the second Dirac points
$E={V_W} +\tau$.
On the other hand, compared to our previous
work~\cite{JRZH14}, we found a strong similarities with respect to
the monolayer case.

\subsection{Graphene superlattices}

 \begin{figure}[h!]
 \centering
 \includegraphics[width=7cm, height=6cm]{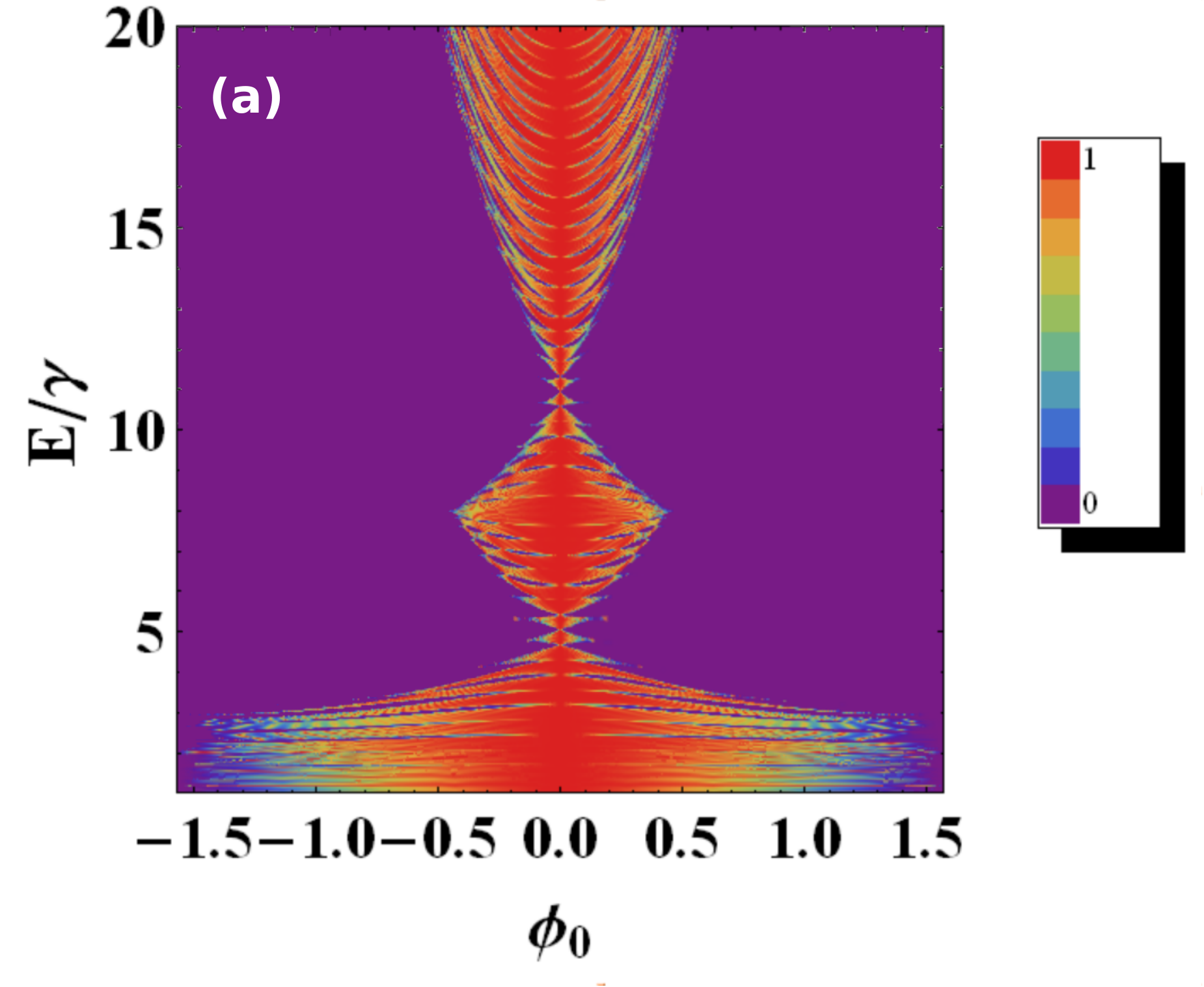}
\quad  \ \ \includegraphics[width=7cm, height=6cm]{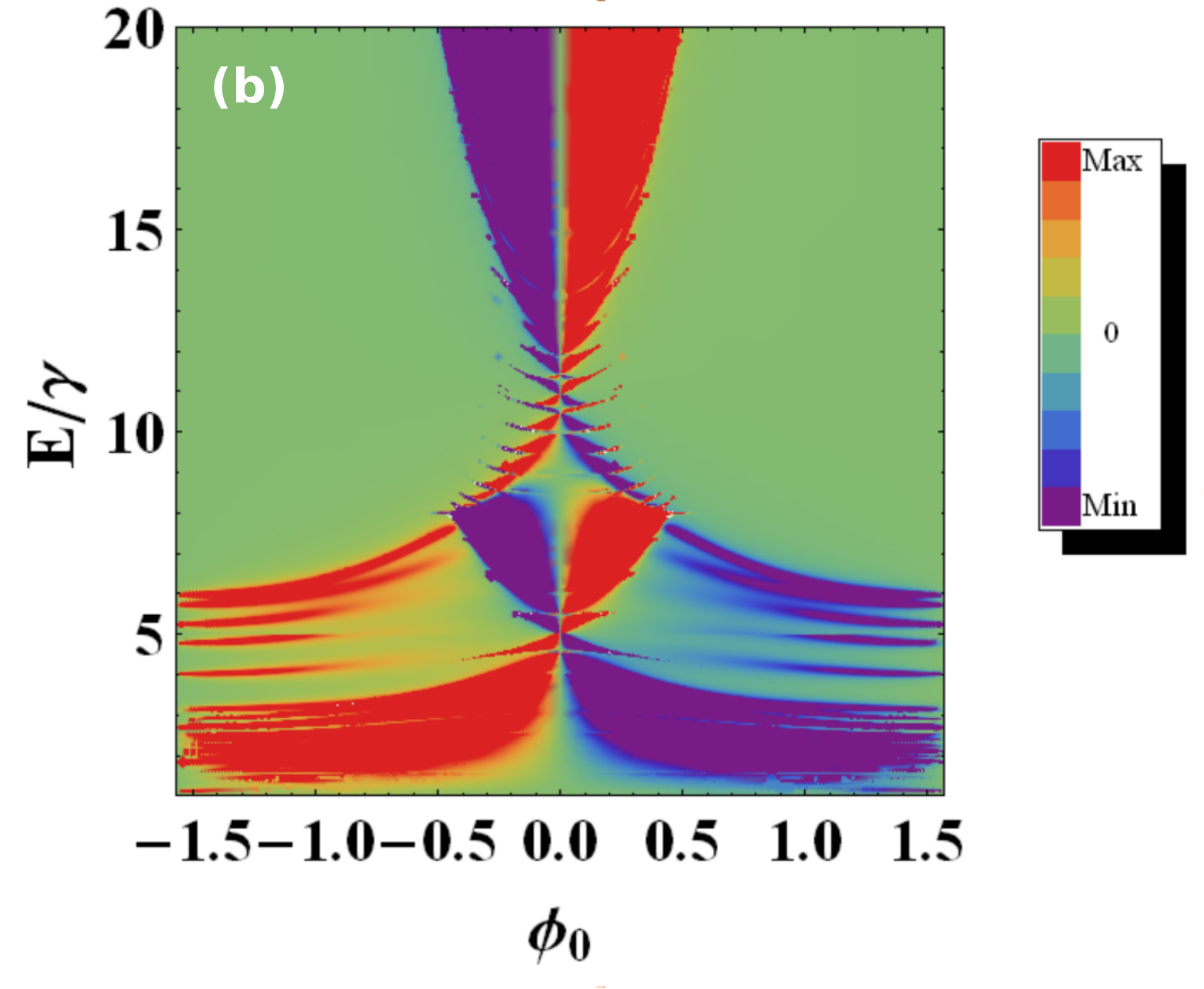}\\
   \includegraphics[width=7cm, height=6cm]{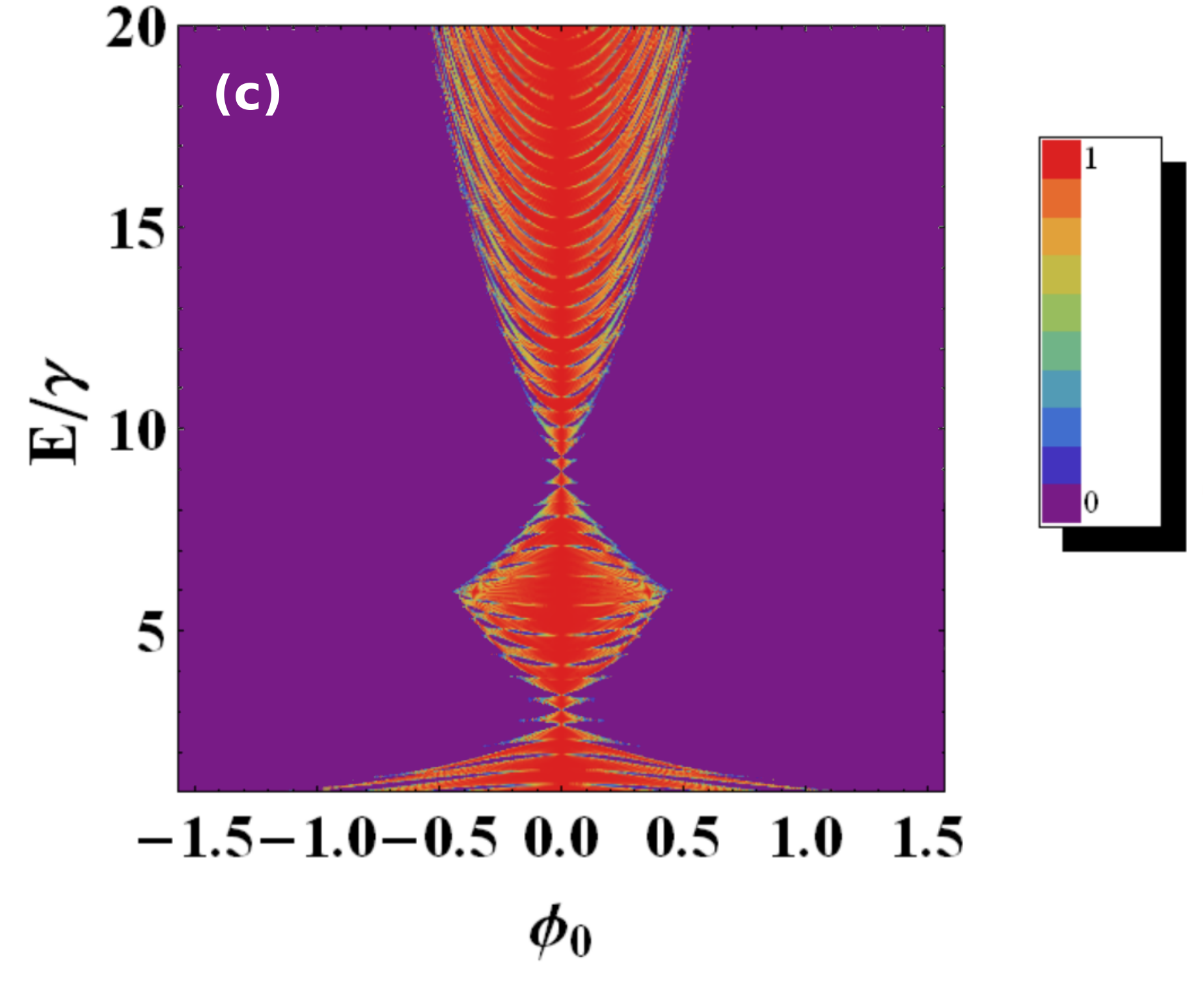}
\quad  \ \ \includegraphics[width=7cm, height=6cm]{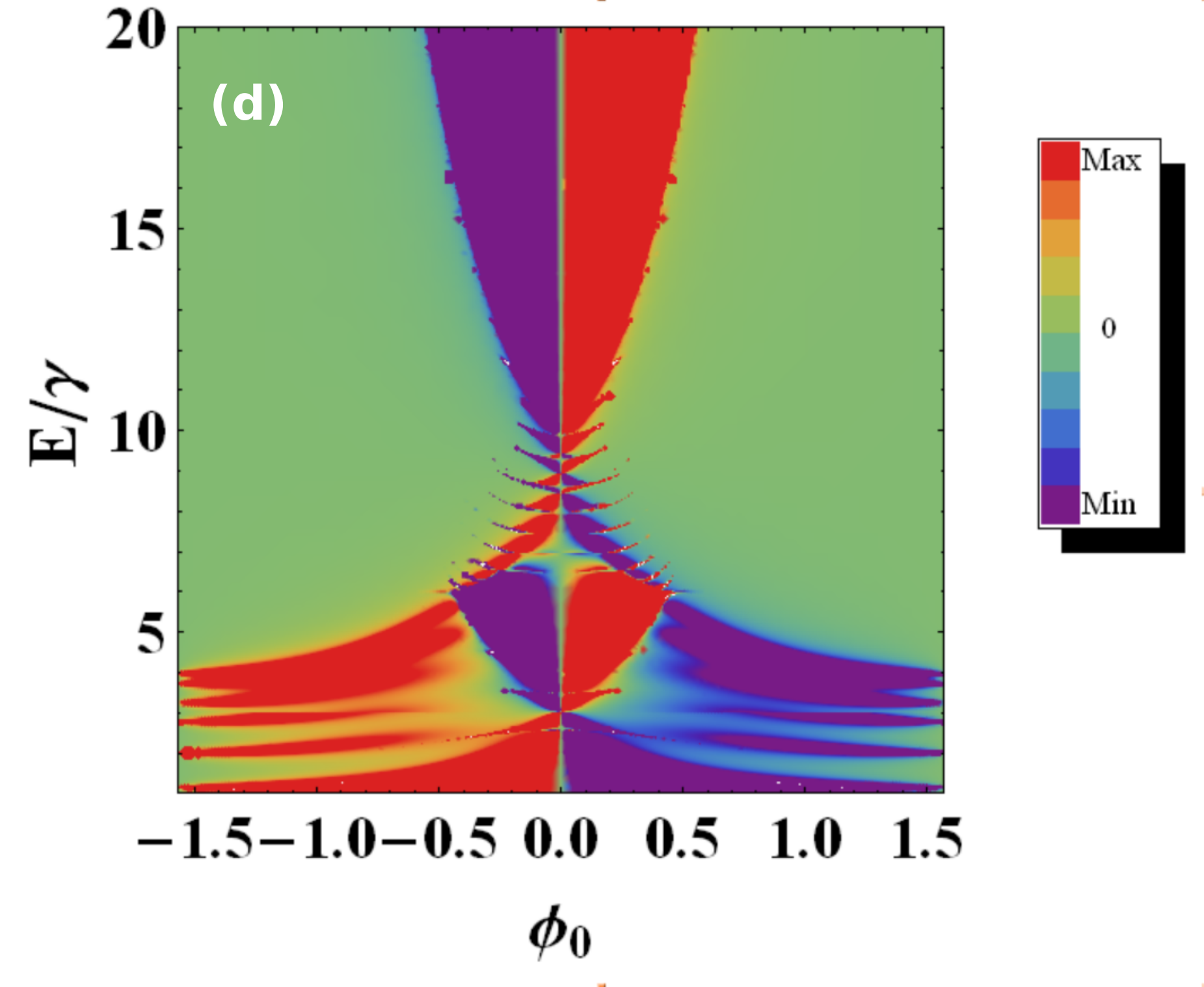}
    \caption{\sf{Density plot of the transmission probability and
Goos-H\"{a}nchen shifts as function of the incident angle and its
energy for graphene superlattices structure ($(BW)^{5}B(WB)^{5}$),
with physical parameters
$\Delta_B=0$, $d_W= 2 d_B =14 nm$, $V_B=10 \gamma$ and $V_W=4
\gamma$.
  $\tau =1$ for (a,b),  $\tau =-1$ for (c,d)
}}\label{TGHS-10P}
\end{figure}

Let us consider a graphene SLs
composed of 11 barriers and 10 wells denoted by
$(BW)^{5}B(WB)^{5}$. To
underline the behaviors of the transmission and GH shifts
in terms of the incident angle and its energy,
we plot
Figure~\ref{TGHS-10P},
with the same physical parameters
as in Figure~\ref{TGHS-2b}. Compared to the results shown for
double barrier structure, we notice that
there are the same
positions of the Dirac points for each individual cone.
For positive incident angle (when $E< V_W + \tau$ and $E> V_B +
\tau$), the shifts are respectively, in the forward and backward
directions, which is due to the fact that the signs of group
velocity are opposite.
We observe that
peaks in transmission gap appear
and the GH shifts display sharp peaks inside the transmission gap
around the Dirac point located at $E=V_W + \tau$,
both of results are absent in the case of double barrier
structures.
One can see that the number of sharp peaks of the shifts is equal
to that of transmission resonances around the two Dirac points.
It is
interesting to note that for
graphene SLs, we
have more then one Dirac point located at the same position,
while the position of the Dirac point is the same whatever the
number of the barriers. Like single and double barrier structures,
the transmission is bilaterally symmetrical with respect to the
normal incidence. In addition, we notice that as observed
in~\cite{JRZH14,AM14},
the GH shifts are
related to the transmission gap around the two Dirac points.\\



 \begin{figure}[h!]
 \centering
 \includegraphics[width=7cm, height=6cm]{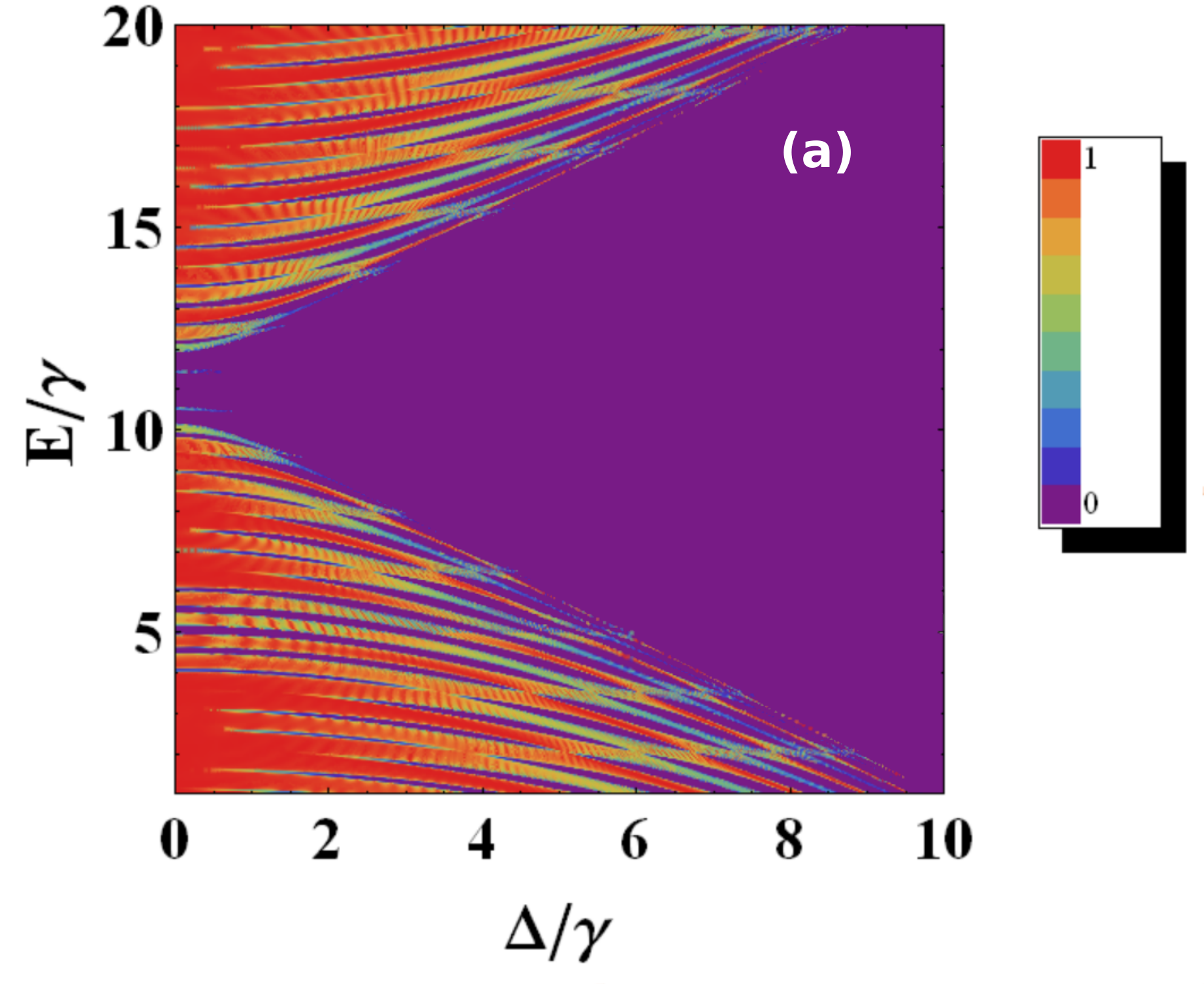}
\quad  \ \ \includegraphics[width=7cm, height=6cm]{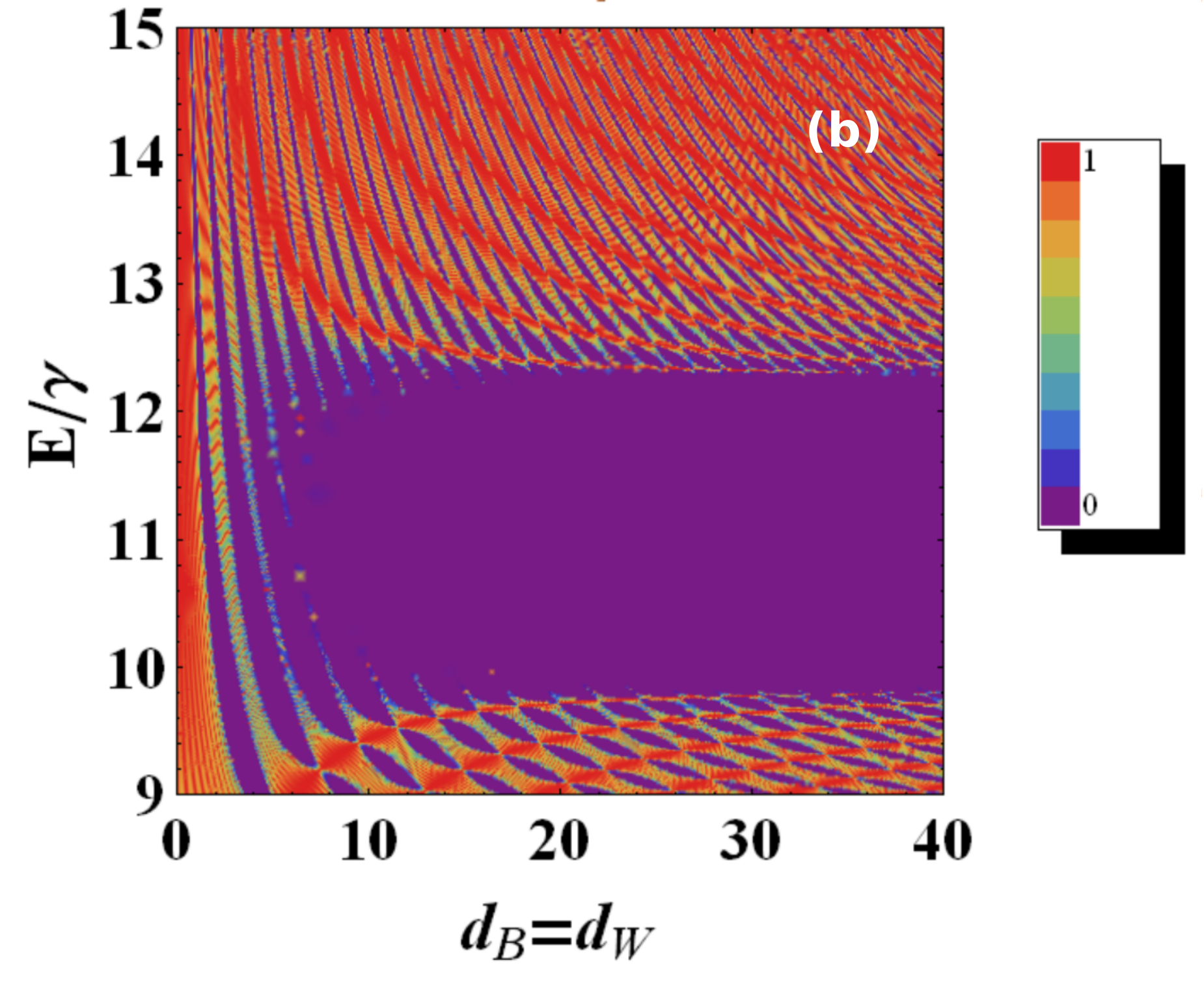}\\
\caption{\sf{Density plot of transmission probability as function
of the gap and the interbarrier $d_B=d_W$ and energy $E$, with
$V_W= 4 \gamma$, $V_B = 10 \gamma$, $\tau = 1$ and
${\phi_{0}^{+1}} = 4^\circ$.
$d_B=d_W= 14 nm$ for (a), $\Delta_B = 1 \gamma$ for
(b).}}\label{TDeltad-10}
\end{figure}

Now, we will turn to
discus the influence of the gap and the width $d_B=d_W$ on the
transmission for graphene SLs. Note that, the gap is introduced,
as shown in Figure~\ref{system}, in the barrier regions. In
Figure~\ref{TDeltad-10}(a), we show the density plot of the
transmission as function of the gap $\Delta_B$ and energy $E$.
This has been performed by fixing the parameters $V_W= 4 \gamma$,
$V_B = 10 \gamma$, $\tau = 1$, $\phi^{+1}_0 = 4^\circ$ and
$d_B=d_W=7 nm$.
For zero gap, one can see that  the transmission exhibits sharp
peaks around the two Dirac points $E=V_B + 1$ and $E=V_W + 1$.
By increasing $\Delta_B$, the transmission gap around the Dirac
point located at $E=V_B + 1$ increases. We also observe that the
transmission
exhibits some oscillation and vanishes after that. It is worth to
see how the barrier width $d_B=d_W$ will affect the transmission
probability, this is shown in Figure~\ref{TDeltad-10}(b).
We
choose the same physical parameters like in
Figure~\ref{TDeltad-10}(a)
with gap
$\Delta_B = 1 \gamma$. As we have
already seen
in Figure~\ref{TDeltad-10}(a), we have a transmission gap around
the Dirac point located at $E=V_B + 1$. In addition, there exists
a sharp transmission peaks and the location of such peaks is
changed by the interbarrier width.


\subsection{Graphene superlattices with defect}

\begin{figure}[h!]
 \centering
 \includegraphics[width=7cm, height=6cm]{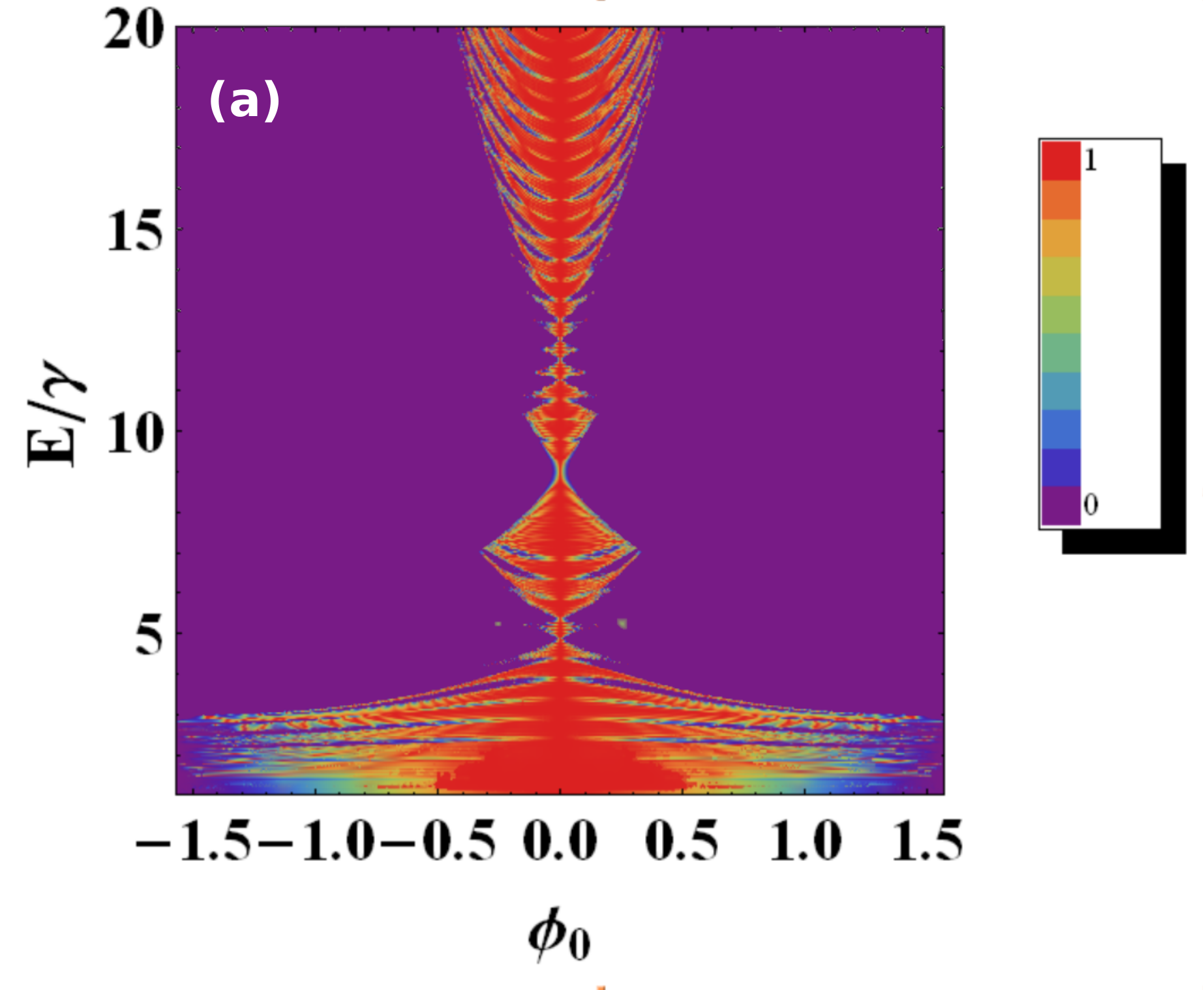}
\quad   \includegraphics[width=7cm, height=6cm]{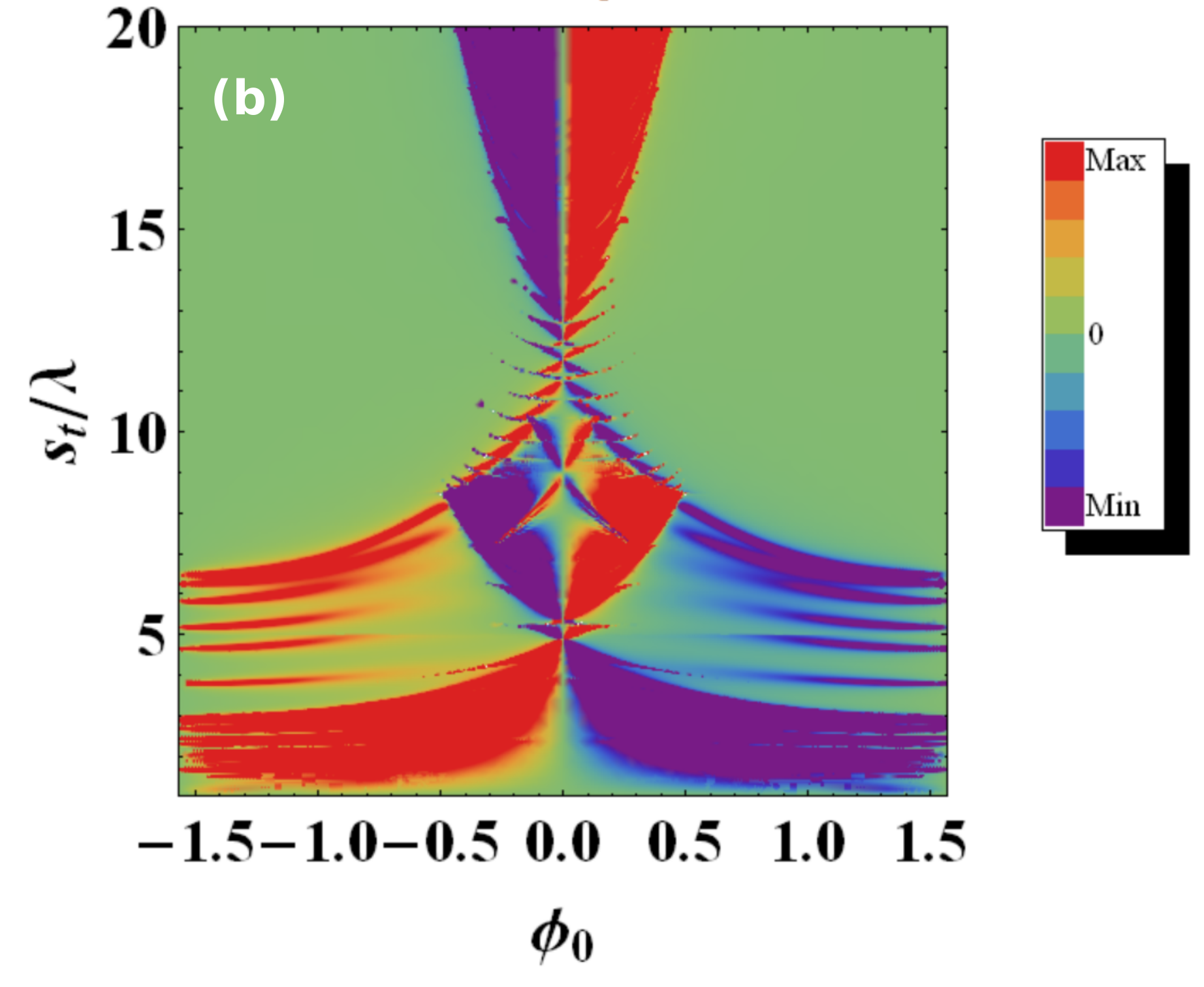}
\caption{\sf{Density plot of the transmission probability (a) and
Goos-H\"{a}nchen shifts (b) as function of the incident angle and
its energy for graphene superlattices $(BW)^{5}D(WB)^{5}$ with
defect $D$.
 $\Delta_B=0$, $\Delta_D=0$, $d_W=2 d_B= 14 nm$, $d_D = 30 nm$, $V_B = 11 \gamma$, $V_W
 = 4 \gamma$ and $V_D = 8 \gamma$.}}\label{TGHS-10D}
\end{figure}

Now, we consider a gapless graphene SLs $(BW)^{5}D(WB)^{5}$ with a
defect $D$.
Figure~\ref{TGHS-10D} presents the numerical results of the
transmission and GH shifts
for the upper cone ($\tau=+ 1$) in terms of the incident angle and
its energy for graphene SLs with defect.
We should emphasize, that for the lower cone ($\tau=-1$) we obtain
the same form as for $T^{+1}$ and $s_t^{+1}$ but just shifted
down.
In such structure, one can clearly end up with an interesting
result such that, in addition to the two Dirac points found in the
case of graphene SLs, it has a third Dirac point located at $E
=V_D + \tau$. Additionally, we have more then one Dirac point
located at the same position for $E=V_B + \tau$ and $E=V_W +
\tau$, but we have only one Dirac point located at $E=V_D + \tau$.
Similar to the case
with defect (Figures~\ref{TGHS-10P}(b,d)), we observe that the
shifts display sharp peaks inside the transmission gap around the
two Dirac points located at $E=V_B + \tau$ and $E=V_W + \tau$,
which are absent in the transmission gap around
$E=V_D + \tau$.

\subsection{Influence of the potential height $V_D$}

 \begin{figure}[h!]
 \centering
 \includegraphics[width=6cm, height=4.5cm]{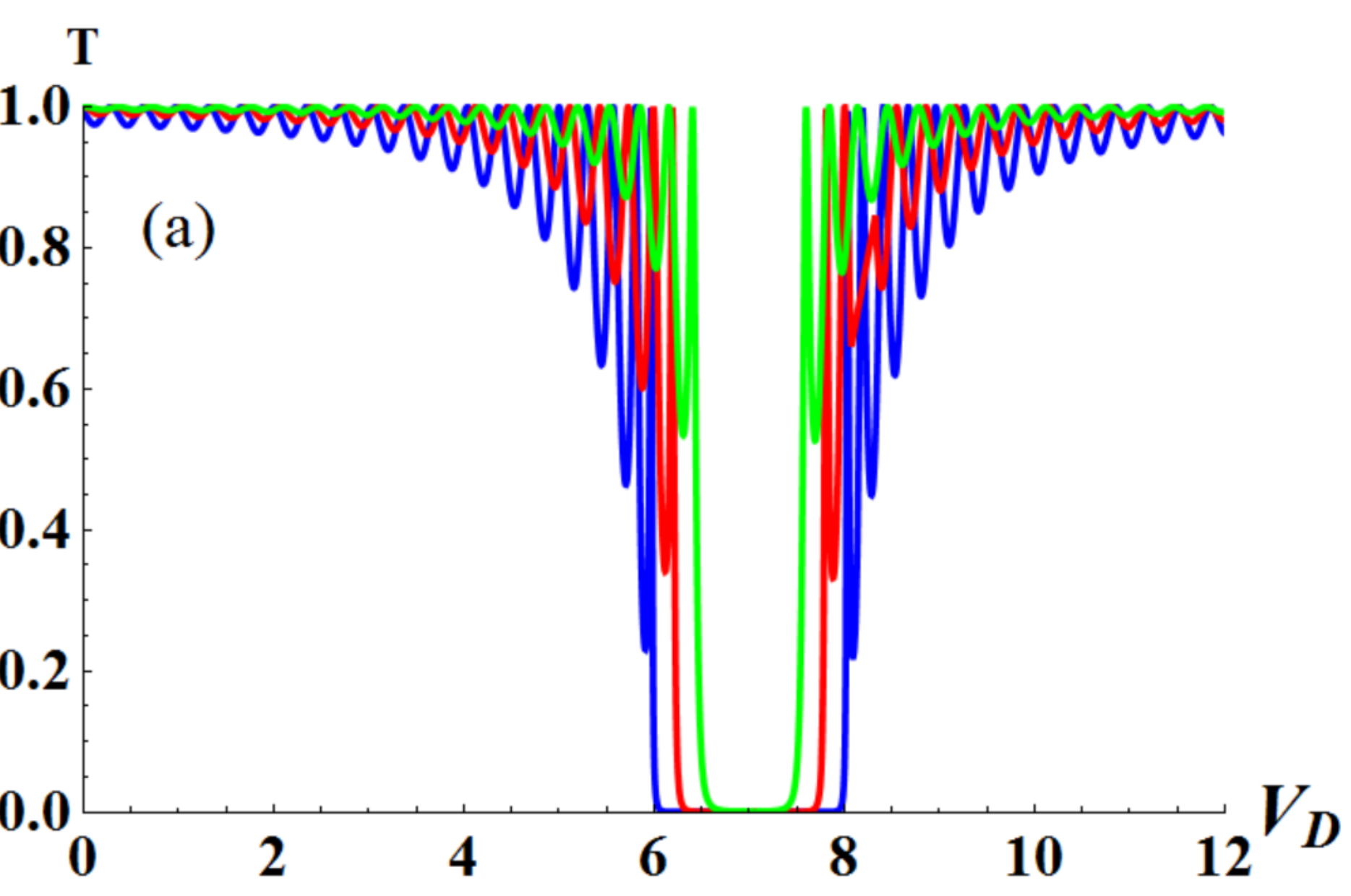}
  \quad \ \ \includegraphics[width=6.5cm, height=4.5cm]{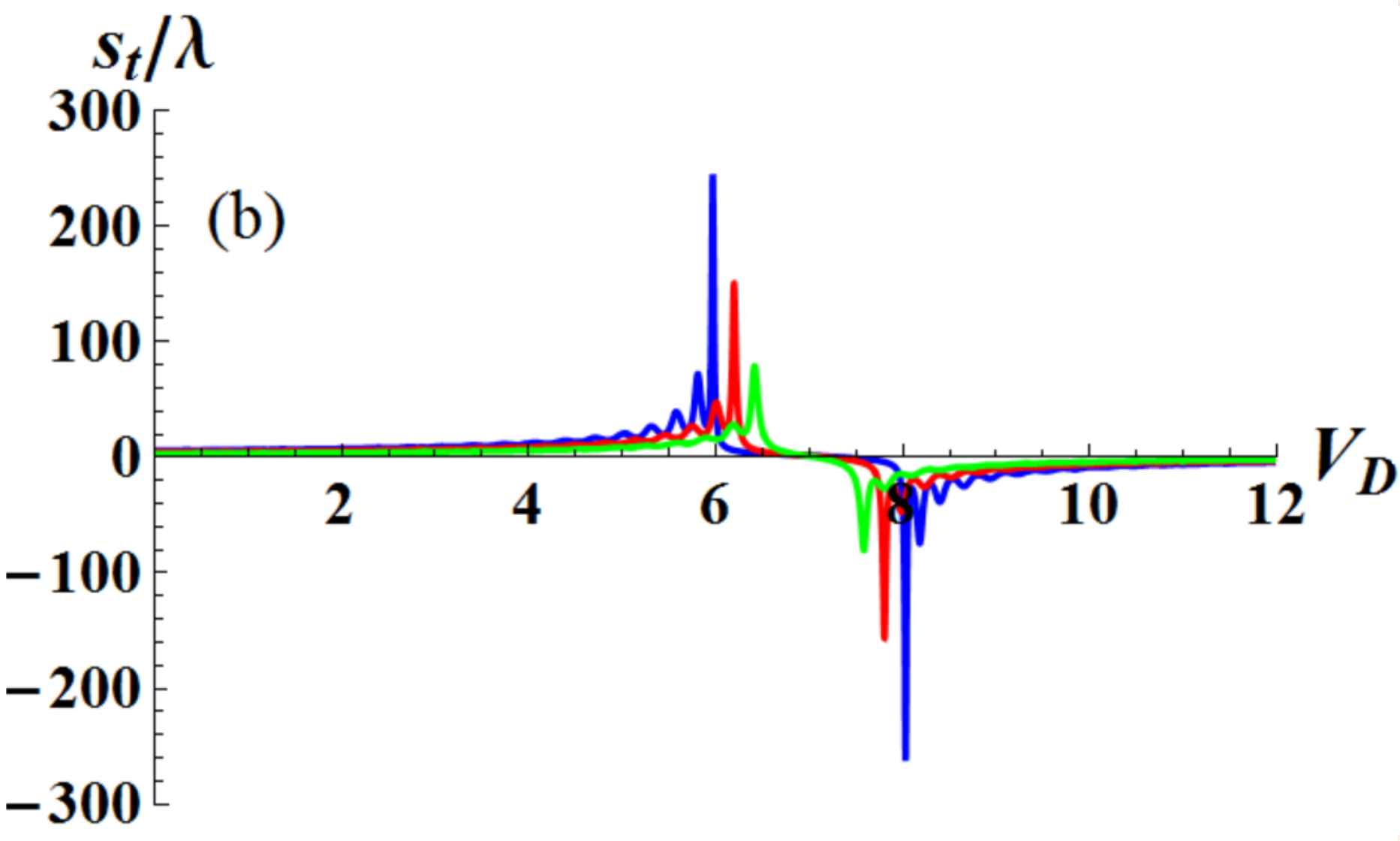}
\caption{\sf{Transmission probability (a) and Goos-H\"{a}nchen
shifts (b) as function of the potential height $V_D$ for graphene
superlattices $(BW)^{5}D(WB)^{5}$ with defect $D$. $\Delta_B=0$,
$d_W= 2 d_B=14 nm$, $d_D = 30 nm$, $V_B = 11 \gamma$, $V_W = 0$,
$\tau = 1$, $\phi_{0}\textcolor[rgb]{1.00,0.00,0.00}{^{+1}}=
4^\circ$ (green line),
$\phi_{0}\textcolor[rgb]{1.00,0.00,0.00}{^{+1}}= 6^\circ$ (red
line), $\phi_{0}\textcolor[rgb]{1.00,0.00,0.00}{^{+1}}= 8^\circ$
(blue line) and $E = 8 \gamma$.}}\label{TGHS-VD}
\end{figure}
Now let us
see how the potential height $V_D$ of defect in gapless graphene
SLs affects the transmission $T^{+1}$
and GH shifts $s_t^{+1}$.
These two quantities are shown in Figure~\ref{TGHS-VD} for $d_W= 2
d_B=14 nm$, $d_D = 30 nm$, $V_B = 11 \gamma$, $V_W = 0$, $\tau =
1$, $E = 8 \gamma$ and three different values of the incident
angle $\phi^{+1}_0= 4^\circ,\ 6^\circ,\ 8^\circ$.
Figure~\ref{TGHS-VD}(b)
shows that the GH shifts change the sign at the Dirac point $E=
V_D + 1$.
One can see that, there is a strong dependence of the GH shifts on
the incident angle. Indeed, by increasing the incident angle
the maximum absolute value of the shifts increase and
the transmission gap becomes larger. We notice that the GH shifts
are positive as long as the condition $ E - \tau > V_D$ is
satisfied and negative if  $ E - \tau < V_D$. For large value of
$V_D$, the GH shifts become mostly constant. We deduce that there
is a strong dependence of the GH shifts on the potential height
$V_D$, which can help to realize controllable negative and
positive GH shifts. From Figure~\ref{TGHS-VD}(a), we clearly see
that the transmission gap becomes larger by increasing the
incident angle. In addition,
by increasing $V_D$ both transmission and GH shifts exhibits an
oscillatory behavior
in terms of the potential $V_D$.

\section{Conclusion}

We
investigated the transmission and Goos-H\"{a}nchen shifts
for the Dirac fermions transmitted through AA-stacked bilayer
graphene superlattices with
a periodic potentials of square barriers.
We
started by formulating our Hamiltonian model that describes the
system under consideration and getting the associated energy
bands. The obtained bands are composed of two Dirac cones shifted
up and down by the interlayer coupling $\gamma$.

Using the transfer matrix method, we
calculated the transmission and the Goos-H\"{a}nchen shifts.
These two quantities were investigated in
different graphene-systems:
single, double barriers and
graphene superlattices. We
obtained two transmissions and two shifts
corresponding to the upper and lower cones. The total transmission
is the average of the two transmissions and the same for the total
Goos-H\"{a}nchen shifts.
Moreover, we found that the two transmissions and the two
Goos-H\"{a}nchen shifts, for both cones, have the same form as
that in the case of monolayer graphene
but shifted up and down by $\tau$.

Subsequently, it has been shown that Klein tunneling of an
electron can occur at normal incidence, for single, double
barriers and graphene superlattices. Also, we have found that the
shifts can be positive as well as negative and change the sign at
the Dirac points. Moreover, by increasing the number of barriers,
the maximum absolute values of the shifts increase.
In the case of double barrier structures and graphene
superlattices,
exist extra Dirac points located at $E=V_W+ \tau$, as compared to
the case of single barrier, where we have only one Dirac point
located at $E=V_B+ \tau$. In addition, for the case of graphene
superlattices,
there is more than one Dirac point at the same position. For such
structure, the Goos-H\"{a}nchen shifts display sharp peaks inside
the transmission gap around the two Dirac point $E=V_B+ \tau$ and
$E=V_W+ \tau$. We ended up with an interesting result such that
the number of sharp peaks is equal to that of transmission
resonances. We noticed that the sharp peaks are absent around the
Dirac point $E=V_W+ \tau$ in the case of double barrier structure.
These results are in agrement with those of monolayer graphene
obtained in our previous work~\cite{JRZH14}.

Furthermore, we showed that the shifts can be modulated by the height of
potential barrier and also can be enhanced by the presence of
resonant energies, which have potential applications in various
graphene-based electronic devices~\cite{WZPX11,ZMC11,SWG12}. In
addition, we also investigated the Goos-H\"{a}nchen shifts
for graphene superlattices with a defect. It is
observed that the negative or positive shifts can be enhanced and
controlled by the potential height of defect.
The Goos-H\"{a}nchen shifts discussed her may have potential
applications in the control of electron beams in the fields of
various graphene based electronic devices.


\section*{Acknowledgment}

The generous support provided by the Saudi Center for Theoretical
Physics (SCTP) is highly appreciated by all authors.

\end{document}